\documentclass[aps,prd,amssymb,eqsecnum,nofootinbib,floatfix,twocolumn, a4paper]{revtex4} 

\usepackage{mathrsfs}
\usepackage{latexsym,amsmath,amsfonts,amssymb} 
\usepackage{bm}
\usepackage{xcolor}

\DeclareMathOperator{\arcsinh}{arcsinh}

\usepackage{graphicx}

\allowdisplaybreaks

\newcommand{\be}{\begin{equation}}
\newcommand{\ee}{\end{equation}}
\newcommand{\beq}{\begin{equation}}
\newcommand{\eeq}{\end{equation}}
\newcommand{\bea}{\begin{eqnarray}}
\newcommand{\eea}{\end{eqnarray}}
\def\nn{\nonumber}

\newcommand{\g}{{\gamma}}
\newcommand{\om}{{\omega}}
\newcommand{\Om}{{\Omega}}
\newcommand{\eps}{\epsilon}
\newcommand{\vareps}{\varepsilon}
\newcommand{\mb}{\bar m}
\newcommand{\pinf}{p_{\infty}}
\newcommand{\ie}{{\it i.e.\,}}

\newcommand{\D}{\partial}

\begin{document}

\title{Comparing One-loop Gravitational Bremsstrahlung Amplitudes to the Multipolar-Post-Minkowskian Waveform}

\author{Donato Bini$^{1,2}$, Thibault Damour$^3$, Andrea Geralico$^1$}
  \affiliation{
$^1$Istituto per le Applicazioni del Calcolo ``M. Picone,'' CNR, I-00185 Rome, Italy\\
$^2$INFN, Sezione di Roma Tre, I-00146 Rome, Italy\\
$^3$Institut des Hautes Etudes Scientifiques, 91440 Bures-sur-Yvette, France
}

\date{\today}

\begin{abstract}
We compare recent one-loop-level, scattering-amplitude-based, computations of the classical part
of the gravitational bremsstrahlung waveform to the frequency-domain version of the corresponding
 Multipolar-Post-Minkowskian
waveform result. When referring the one-loop result to the classical averaged momenta $\bar p_a = \frac12 (p_a+p'_a)$,
the two waveforms are found to agree at the Newtonian and first post-Newtonian levels,
as well as at the first-and-a-half post-Newtonian level, i.e. for the leading-order quadrupolar tail.
However, we find that there are significant differences at the second-and-a-half post-Newtonian level,
$O\left( \frac{G^2}{c^5} \right)$, i.e.
when reaching: (i) the first post-Newtonian correction to the linear quadrupole tail; (ii) Newtonian-level linear tails
of higher multipolarity (odd octupole and even hexadecapole); (iii) radiation-reaction effects on the worldlines;
and (iv) various  contributions of cubically nonlinear origin (notably  linked to
the quadrupole$\times$ quadrupole$\times$ quadrupole coupling in the wavezone).
These differences are reflected at the sub-sub-sub-leading level in the soft expansion, $  \sim \om \ln \om $, i.e. $O\left(\frac{1}{t^2} \right)$
in the time domain.
Finally, we computed the first four terms of the low-frequency expansion of the Multipolar-Post-Minkowskian waveform and checked
that they agree with the corresponding existing classical soft graviton results.
\end{abstract}

\maketitle


\section{Introduction}

The last years have witnessed a fruitful dialogue between traditional analytical approaches  to the dynamics
and radiation of gravitationally interacting binary systems (based
on perturbatively solving the classical Einstein equations), and quantum-based approaches
capitalizing on advances in quantum field theory and in the computation of scattering amplitudes.
This useful classical-quantum synergy was first developed in the context of the computation of 
higher post-Minkowskian (PM) contributions to the scattering
of two massive particles (with or without spin)\cite{Adamo:2022ooq,Amati:1990xe,Aoude:2020onz,Bern:2019nnu,Bern:2019crd,Bern:2020gjj,Bern:2021yeh,Bjerrum-Bohr:2018xdl,Bjerrum-Bohr:2019kec,Bjerrum-Bohr:2021vuf,Bjerrum-Bohr:2021din,Bjerrum-Bohr:2021wwt,Bjerrum-Bohr:2022ows,Bonocore:2021qxh,Brandhuber:2021eyq,Cheung:2018wkq,Cristofoli:2019neg,Cristofoli:2020uzm,Damgaard:2019lfh,Damgaard:2021ipf,Damgaard:2023vnx,Damour:2016gwp,Damour:2017zjx,Damour:2019lcq,Damour:2020tta,DiVecchia:2019myk,DiVecchia:2019kta,DiVecchia:2020ymx,DiVecchia:2021ndb,DiVecchia:2021bdo,DiVecchia:2022owy,DiVecchia:2022piu,DiVecchia:2023frv,Dlapa:2021vgp,Dlapa:2022lmu,Dlapa:2021npj,Dlapa:2023hsl,FebresCordero:2022jts,Herrmann:2021tct,Jakobsen:2021zvh,Jakobsen:2022zsx,Jakobsen:2022fcj,Jakobsen:2022psy,Jakobsen:2023ndj,Kalin:2019rwq,Kalin:2020mvi,Kalin:2020fhe,Kalin:2020lmz,Kalin:2022hph,KoemansCollado:2019ggb,KoemansCollado:2018hss,Kosower:2018adc,Maybee:2019jus,Liu:2021zxr,Mogull:2020sak,Parra-Martinez:2020dzs}.
More recently, various flavours
of Effective Field Theory (EFT) were applied to the computation of the gravitational wave (GW) emission
during high impact parameter collisions of binary systems \cite{Luna:2017dtq,Mougiakakos:2021ckm,Jakobsen:2021lvp,Cristofoli:2021jas,Herrmann:2021lqe,Jakobsen:2021smu,Cristofoli:2021vyo,Riva:2022fru,Heissenberg:2022tsn,Manohar:2022dea,Adamo:2022qci,Elkhidir:2023dco,Brandhuber:2023hhy,Herderschee:2023fxh,Georgoudis:2023lgf,Gonzo:2023cnv}.

The gravitational wave (GW) emission from gravitationally interacting binary systems has been intensely studied 
by classical approximation methods for many years,
especially since the 1980's when arose the motivation to help the future detection, and analysis, of GW signals by 
the then developing network of ground-based GW detectors.
One of the most successful {\it analytical} method for computing the GW emission from generic sources has been the 
Multipolar-Post-Minkowskian (MPM) formalism \cite{Blanchet:1985sp,Blanchet:1986dk,Blanchet:1987wq,Blanchet:1989ki,Damour:1990gj,Damour:1990ji,Blanchet:1992br,Blanchet:1995fr,Poujade:2001ie}.
The MPM formalism has been developed over the years to a high perturbative accuracy, recently reaching  the 
fourth post-Newtonian (4PN) accuracy \cite{Blanchet:2023bwj,Blanchet:2023sbv}, \ie the N$^4$LO level in a PN expansion beyond the leading-order 
quadrupole formula pioneered by Einstein \cite{Einstein:1918btx}. 
[Note that this PN accuracy includes nonlinear terms of fractional 
fourth PM order ($G^4$) beyond the leading-order (``Newtonian-level") quadrupole formula.]

The aim of the present paper is to compare the MPM waveform to recently derived expressions of the (frequency-domain)
waveform obtained by quantum-based EFT computations. The most recent computations have derived the
classically-relevant part of the one-loop scattering amplitude with four scalars and one graviton, and its corresponding
classical waveform \cite{Brandhuber:2023hhy,Herderschee:2023fxh,Georgoudis:2023lgf}.

The quantum one-loop accuracy corresponds to computing the classical waveform~\footnote{We use the mostly plus signature,
 $\eta_{\mu \nu}={\rm diag}(-1,+1,+1,+1)$.}
 $h_{\mu \nu} = g_{\mu \nu}-\eta_{\mu \nu}$ to order $G^3$, i.e.
\bea 
\label{hmunu}
h_{\mu \nu}&=&G h^{\rm lin}_{\mu \nu}+ G^2 h^{\rm post-linear \, or\, tree}_{\mu \nu}\nonumber\\
&+& G^3 h^{\rm post-post-linear \, or\,  one-loop}_{\mu \nu} +O(G^4)\,,
\eea
where we indicated the dictionary between the quantum nomenclature (tree, one-loop,...) and the classical PM one
(linear in $G$, post-linear $\sim G^2$, post-post-linear $\sim G^3$,...). Let us immediately mention two
somehow confusing facts concerning the $G$ structure of the waveform Eq. \eqref{hmunu}. First,
when studying the frequency-domain waveform in a PM-expanded way, the linear contribution $G h^{\rm lin}_{\mu \nu}$,
being ``Coulombic," {\it i.e.} purely stationary in the time-domain, is localized at zero frequency. [It does not even appear
in (some of) the quantum amplitude computations.] Second, when considering the combined multipolar-PM-PN expansion of the
waveform, as recorded at infinity, it is useful to think in terms of {\it fractional} corrections to the leading-order quadrupole
waveform, as described by time-derivatives of the quadrupole moment of the system. 
The wish to work with a waveform quantity expressible in terms of the Newtonian quadrupole in the non-relativistic
limit naturally leads one to define the following $ \frac{c^4}{4 G}$-rescaled~\footnote{We indicate powers of $c$ when they help to locate which quantities have a non-relativistic limit. Otherwise, we often set $c$ to 1 without warning.}, frequency-space complex waveform (involving a complex null
polarization vector $ \eps^{\mu }$ orthogonal to the momentum $k^\mu=\om (1,{\bf n})$ of the emitted graviton)
\bea
W(\om,  \theta,\phi) &\equiv& \frac{c^4}{4 G} \lim_{R\to \infty}(R( h_+ -  i h_\times))\nonumber\\
&=&\frac{c^4}{4 G}\lim_{R\to \infty} \eps^{\mu } \eps^{\nu }R\, h_{\mu \nu}\,.
\eea
As we shall discuss in detail below, in the time-domain, the leading-post-Newtonian contribution to $W$ is the 
classic Einstein quadrupole
result $W^{\rm Newtonian}(t)= \frac12 \eps^i \eps^j \frac{d^2}{dt^2} Q_{ij}$, where~\footnote{The symbol STF
denotes a Cartesian symmetric-trace-free projection.} $Q_{ij}=(\sum_a m_a x_a^ix_a^j)^{\rm STF}$. Though the time-domain
expression of $W^{\rm Newtonian}$ starts at order $G^0$ (because of the presence of the term
$2  \,(m_1 v_1^i v_1^j+  m_2 v_2^i v_2^j)^{\rm STF}$ in $\frac{d^2}{dt^2} Q_{ij}$), its time-dependence starts at $O(G^1) $.
 Indeed, the third derivative of the Newtonian quadrupole $Q_{ij}$ vanishes along the asymptotic straight-line motions.
As a consequence, only the zero-frequency part of $W^{\rm Newtonian}(\om,  \theta,\phi)$  is of order $G^0$ (and given by the asymptotic values of
the velocity tensor $2  \, \lim_{t\to\pm\infty} \,(m_1 v_1^i v_1^j+  m_2 v_2^i v_2^j)^{\rm STF}$), while
its non-zero-frequency part starts at order $G^1$.
The $G$ expansion (for non-zero frequencies) of $W(\om,  \theta,\phi)$ then reads
\bea 
\label{Woneloop}
W(\om,  \theta,\phi) &=& G W^{\rm post-linear \, or\, tree} \nonumber\\
&+& G^2 W^{\rm post-post-linear \, or\,  one-loop}\nonumber\\
&+& O(G^3)\,,
\eea
where we indicated the classical (and quantum) origin of each order in $G$.

The leading-PM-order (tree-level) contribution to $W$ was first computed in the time-domain by Peters \cite{Peters:1970mx}  and 
Kovacs and Thorne \cite{Kovacs:1977uw,Kovacs:1978eu}
using classical perturbation theory. It has been recently recomputed in an efficient manner
using quantum-based (or quantum-related) EFT approaches by several groups 
\cite{Mougiakakos:2021ckm,DiVecchia:2021bdo,Jakobsen:2021smu} 
(see also the double-copy computations of  Refs. \cite{Goldberger:2017frp,Luna:2017dtq}).
 Recently three different groups independently computed the one-loop
contribution to the waveform, and got results in agreement with each other \cite{Brandhuber:2023hhy,Herderschee:2023fxh,Georgoudis:2023lgf}. 

In order to explain the method we used to compare the one-loop-accurate EFT-based waveform, Eq. \eqref{Woneloop}, derived
in Refs. \cite{Brandhuber:2023hhy,Herderschee:2023fxh,Georgoudis:2023lgf}
with the (time-domain) results of the MPM formalism we shall start
by briefly summarizing the basic features of both computations.

\section{Brief reminder of the Multipolar-Post-Minkowskian (MPM) formalism}

 The MPM formalism solves Einstein's equations by combining several analytical tools: In the exterior zone of the
 binary system, it solves the vacuum Einstein equations by combining a PM expansion of the metric, schematically
 $g=\eta + G h^{\rm 1PM} + G^2 h^{\rm 2PM} + G^3 h^{\rm 3PM} +\cdots$, with multipolar
 expansions, starting at the linear level: $h^{\rm 1PM}=\sum_{\ell^{+}} h^{\rm 1PM}_{ \ell^+ , {\rm ret}}
 + \sum_{\ell^{-}} h^{\rm 1PM}_{ \ell^- , {\rm ret}}$, where $ h^{\rm 1PM}_{ \ell^\pm , {\rm ret}}$ 
 denotes the most general {\it retarded} ({\it i.e.} outgoing) multipolar wave of multipole order $\ell$ and
 spatial parity $\pm$. Einstein's equations are then  iteratively solved (in the exterior zone)
 by using {\it retarded propagators} at each
 PM nonlinear order, e.g., $ h^{\rm 2PM} = {\rm FP}_B \Box^{-1}_{\rm ret} \left( \left(\frac{|{\bf x}|}{ r_0}\right)^B \D \D  h^{\rm 1PM}  h^{\rm 1PM}\right)$, where FP denotes the finite part of the analytical continuation of the retarded
 integral at $B=0$. [It cures the formal ultra-violet divergences occurring at $r=|{\bf x}| \to 0$ in the exterior MPM iteration scheme.]
  In the near zone of the system, it solves the inhomogeneous Einstein equations by 
 using PN expansions.  One then transfers information between the two zones by using the method
 of matched asymptotic expansions (see Refs. \cite{Blanchet:1987wq,Blanchet:1989ki,Damour:1990ji,Blanchet:1995fr,Poujade:2001ie}). For instance, this method allowed one to find that the first
 non-local-in-time (hereditary) contribution to the near-zone physics arose at the 4PN, and 4PM ($\frac{G^4}{c^8}$)
 level \cite{Blanchet:1987wq}. It also allowed one to derive many types of hereditary contributions to the emitted
 waveform \cite{Blanchet:1992br} (For an updated account see Ref. \cite{Blanchet:2013haa}).

 When extracting the radiation emitted at infinity by the system, the MPM formalism changes from the source-based
 harmonic coordinate system $(x^\mu)$ to some asymptotic Bondi-like coordinate system $T_r, R, \theta,\phi$ \cite{Blanchet:1986dk}.
 Here  $T_r \simeq t- \frac{r}{c} - 2 \frac{G {\cal M}}{c^3} \ln \frac{r}{c b_0}$ denotes a retarded time which contains,
  in particular, a logarithmic shift proportional to the total mass-energy of the system, $ {\cal M} \equiv \frac{E}{c^2}$.
  The definition of $T_r$ involves some arbitrary time scale $ b_0$  (which can be taken to be different from the length scale $r_0$ 
 entering the MPM exterior iteration scheme).
 The MPM formalism is usually set up in the (incoming) center-of-mass (c.m.) of the system. We shall
 denote the complex null frame tangent to the sphere at infinity, {\it i.e.} 
 orthogonal to the unit vector ${\bf n}(\theta,\phi) = (\sin \theta \cos \phi, \sin \theta \sin \phi,  \cos \theta)$ 
recording the direction of emission of the GW as  ${\bf m}$,  $\bar {\bf m}$. We define
\bea 
{\bf m}&=& \frac{1}{\sqrt{2}} \left(\D_\theta {\bf n}(\theta,\phi) + \frac{i}{\sin \theta} \D_\phi {\bf n}(\theta,\phi)\right)\,,\nn\\
\bar {\bf m}&=& \frac{1}{\sqrt{2}} \left(\D_\theta {\bf n}(\theta,\phi) - \frac{i}{\sin \theta} \D_\phi {\bf n}(\theta,\phi)\right)\,,
\eea
and use $\bar {\bf m}$ as null polarization vector (corresponding to $\eps^\mu$ in the one-loop computations).
 
The (transverse-traceless) asymptotic waveform  $\lim_{R\to \infty}( R \, h_{ij}^{\rm TT})$ is then conveniently
recorded in the complex waveform
\bea
h_c(T_r,\theta,\phi) &=&\lim_{R\to \infty}(R( h_+ -  i h_\times))\nonumber\\
&=&\lim_{R\to \infty} \mb^{\mu } \mb^{\nu }R\, h_{\mu \nu}\,.
\eea
The MPM formalism computes each term of the {\it multipolar decomposition} of $h_c(T_r,\theta,\phi)$
in irreducible representations of the rotation group. Namely
 (using $\eta \equiv \frac1c$ as PN expansion parameter)
 \be
 h_c(T_r,\theta,\phi) \equiv 4 G \eta^4  W(T_r,\theta,\phi)\,,
 \ee
 with
\bea
W^{\rm MPM}(T_r,\theta,\phi)&=& U_2+ \eta (V_2 +U_3) + \eta^2 (V_3+U_4)\nonumber\\ 
&+& \eta^3 (V_4+U_5)+ \cdots \,.
\eea
Here, each $U_\ell$ (respectively $V_\ell$) denotes an even-parity (resp. odd-parity) $2^\ell$ radiative multipole contribution, 
expressed in terms
of symmetric-trace-free (STF) Cartesian tensors of order $\ell$, normalized according to
\bea
U_\ell(T_r,\theta,\phi) &=& \frac{1}{\ell!} \mb^{i} \mb^{j } n^{i_1} n^{i_2} \cdots n^{i_{\ell-2}} U_{i j i_1 i_2 \cdots i_{\ell-2}}(T_r)\,, \nn\\
V_\ell(T_r,\theta,\phi) &=& - \frac{1}{\ell!}\frac{2\ell}{\ell+1}\mb^{i} \mb^{j } n^c  n^{i_1} n^{i_2} \cdots n^{i_{\ell-2}}\times \nonumber\\
&&  \epsilon_{cd i}V_{j d i_1 i_2 \cdots i_{\ell-2}}(T_r)\,.
\eea
The (post-Newtonian-matched) MPM formalism relates each (time-domain~\footnote{For simplicity we henceforth denote the
retarded time variable simply as $t$.}) {\it radiative multipole moment} 
$U_{i_1 i_2 \cdots i_{\ell}}(t)$,  $V_{i_1 i_2 \cdots i_{\ell}}(t)$ (observed at future null infinity)
to the source in several steps. 
In a first step, $U_{i_1 i_2 \cdots i_{\ell}}(t)$ and $V_{i_1 i_2 \cdots i_{\ell}}(t)$ are computed
by iteratively solving  Einstein's vacuum equations in the exterior zone,
as a Post-Minkowskian (PM) expansion in terms of two sequences of (intermediate) ``canonical" 
(mass-type and spin-type) moments, $M_{i_1 i_2 \cdots i_{\ell}}(t)$,  $S_{i_1 i_2 \cdots i_{\ell}}(t)$.
In a second step, the latter canonical moments are computed as a PM expansion in terms of 
six sequences of {\it source multipole moments}. These comprise  two main source moments, of mass-type,  $I_{i_1 i_2 \cdots i_{\ell}}(t)$,  and spin-type, $J_{i_1 i_2 \cdots i_{\ell}}(t)$,  together with four auxiliary gauge-type moments, $W_{i_1 i_2 \cdots i_{\ell}}(t)$,  $X_{i_1 i_2 \cdots i_{\ell}}(t)$, $Y_{i_1 i_2 \cdots i_{\ell}}(t)$,  $Z_{i_1 i_2 \cdots i_{\ell}}(t)$. [The latter auxiliary
moments  parametrize a coordinate transformation that is used, at the first iteration of the MPM formalism,
to transform the multipole expansion of the metric outside the source in a form close to the
canonical one, parametrized only by $M_{i_1 i_2 \cdots i_{\ell}}(t)$ and  $S_{i_1 i_2 \cdots i_{\ell}}(t)$ 
(see \cite{Blanchet:2013haa} for details). [Only the monopole component 
$W(t)= [W_{i_1 i_2 \cdots i_{\ell}}(t)]^{\ell=0}$ of the first auxiliary moment 
(which parametrizes a shift of the time coordinate) will play a role below.]   
Finally, using a post-Newtonian-based (PN-based) matching of asymptotic expansions, the source moments
are expressed (in a PN-expanded way)  in terms of the dynamical variables of the system.
Combining the various steps ultimately leads to expressing the radiative moments
as a sum of (linear and nonlinear) retarded functionals  of the (main and auxiliary) source moments,
given as PN-expanded expressions in terms of the dynamical variables of the system. See below.

Fourier-transforming (over the retarded time variable) the radiative moments,
\bea
U_{i_1 i_2 \cdots i_{\ell}}(\om)= \int_{- \infty}^{+ \infty} dt e^{i \om t} U_{i_1 i_2 \cdots i_{\ell}}(t)\,, \nn \\
V_{i_1 i_2 \cdots i_{\ell}}(\om)= \int_{- \infty}^{+ \infty} dt e^{i \om t} V_{i_1 i_2 \cdots i_{\ell}}(t)\,,
\eea
then leads to a multipole expansion of the frequency-domain complex waveform 
$h_c(\om,\theta,\phi) \equiv 4 G \eta^4 W(\om,  \theta,\phi)$ of the form
\bea
\label{Wom_th_phi}
W^{\rm MPM}(\om,  \theta,\phi) &\equiv& U_2(\om,\theta,\phi)\nonumber\\
&+& \eta (V_2(\om,\theta,\phi) +U_3(\om,\theta,\phi))\nonumber\\ 
&+& \eta^2 (V_3(\om,\theta,\phi)+U_4(\om,\theta,\phi)) + \cdots\nonumber\\ 
\eea
For instance the quadrupole contribution reads
\be
U_2(\om,\theta,\phi)=  \frac{1}{2!} \mb^{i} \mb^{j } U_{i j}(\om)\,,
\ee
where the radiative quadrupole moment $ U_{i j}$ differs from the leading-PN-order text-book quadrupole
formula $ U^{\eta^0}_{i j}= \frac{d^2}{dt^2} Q_{ij}$ by many additional, nonlinear PN and PM contributions that
have been successively derived in the MPM formalism (up to the recent 4PN+ 4.5PN accuracy \cite{Blanchet:2023bwj,Blanchet:2023sbv}).

Of particular importance in the following will be the contributions to the multipole moments that are {\it time-dissymmetric}. The
time-dissymmetry can come either directly from radiation-reaction contributions to the (time-domain) multipole moments,
or from hereditary effects coming from retarded graviton propagators in the wave zone.

Several types of time-dissymmetric (hereditary) effects arise in the GW emission, such as (using the
terminology of Ref. \cite{Blanchet:2013haa}): tails, tails of tails, nonlinear interactions, memory, etc. These contributions (which are
generally hereditary) have been evaluated
(in the time domain) to high PM and PN accuracy within the MPM formalism \cite{Blanchet:1992br,Blanchet:1997jj,Blanchet:2013haa}. 
In addition, one must take into account
(when computing the frequency-domain waveform) the effect of radiation-reaction on the scattering motion. 

\begin{figure}
\includegraphics[scale=0.40]{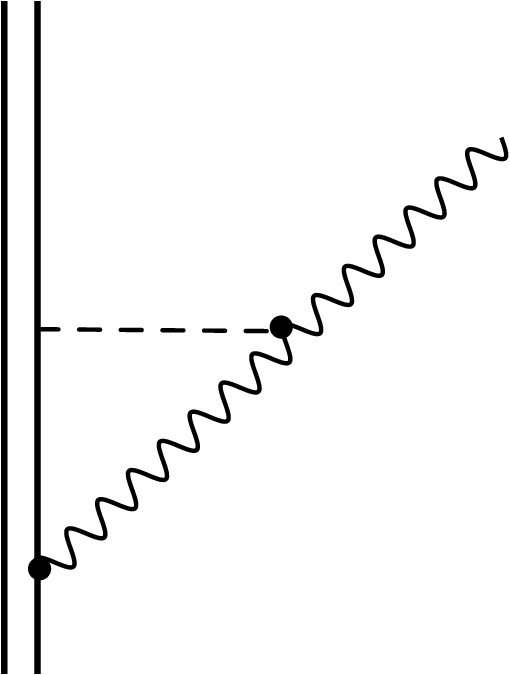}
\caption{\label{fig:1} Diagram illustrating the origin of tail contributions to the waveform, namely the coupling between
a (retarded) radiative multipolar wave (of various multipolarities $\ell^{\pm}$) and a stationary contribution to the
metric ($\frac{G \cal M}{r}$ for the usual linear tail, e.g. Eq. \eqref{U2tail}, or the angular-momentum-generated term 
$\frac{G  L}{r^2}$ for the $U_{ij}^{\rm LQ}$ term, Eq. \eqref{U2LQ}, or its  $V_{ijk}^{\rm LQ}$ and $U_{ijkl}^{\rm LQ}$
analogs). The two wiggly propagators here are retarded.}
\end{figure}

At the
fractional 2.5PN accuracy at which we work here, we must include the following contributions in the 
radiative quadrupole moment $U_{ij}$ (denoting $f^{(n)}(t) \equiv\frac{ d^n}{dt^n} f(t)$):
\bea
\label{U_ij_gen_all}
U_{ij}(t) &=&\left[ I_{ij}^{(2)}\right]^{\leq 2.5PN} + U^{\rm tail}_{ij}(t) +  U^{QQ}_{ij}(t)\nonumber\\ 
&+&  U^{LQ}_{ij}(t) +  U^{WQ}_{ij}(t) \;\; (+  U^{\rm memory}_{ij}(t))
\,.
\eea
Here: $ U^{\rm tail}_{ij}(t)$ denotes the  linear tail contribution  to the radiative quadrupole moment $U_{ij}$ (which starts at fractional order $\eta^3$
 and must be computed at accuracy $\eta^5$). It is given by (denoting ${\mathcal M}\equiv \frac{E}{c^2}$,  where $E$ is the total c.m. energy of the system)
 \bea 
\label{U2tail}
 U_{ij}^{\rm tail}(t)&=& \frac{2G {\mathcal M}}{c^3}\int_0^\infty d\tau I_{ij}^{(4)}(t-\tau) \left(\ln\left(\frac{\tau}{2 b_0}\right)+ \frac{11}{12} \right).\qquad
 \eea
 The physics behind this tail term (and its higher-multipolar analogs discussed below) is illustrated in Fig. \ref{fig:1}.

The QQ term $ U^{QQ}_{ij}(t) $ denotes a nonlinear contribution arising from a cubic quadrupole $\times$ quadrupole $\times$ quadrupole coupling in the action \cite{Blanchet:1997ji}. It reads\footnote{As a rule we call QQ-term the contribution $U_{ij}^{QQ}$ to $U_{ij}$, LQ term the contribution $U_{ij}^{LQ}$ to $U_{ij}$, etc.}
   \bea \label{U2QQ}
U_{ij}^{\rm QQ}(t)&=&\frac{G}{c^5}\left(\frac17 I_{a\langle i}^{(5)}I_{j\rangle a}-\frac57 I_{a\langle i}^{(4)}I_{j\rangle a}^{(1)}-\frac27 I_{a\langle i}^{(3)}I_{j\rangle a}^{(2)}\right),\qquad
\eea
where the angular brackets denote a STF projection: 
$
A^{\langle i_1 i_2\ldots  \rangle}={\rm STF}_{i_1 i_2\ldots }[A^{i_1 i_2\ldots }]
$.

\begin{figure}
\includegraphics[scale=0.4]{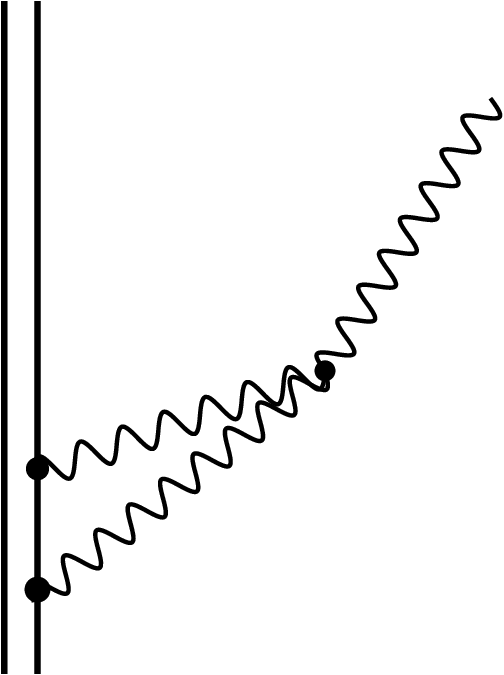}
\caption{\label{fig:2} Diagram illustrating the cubically nonlinear origin of the various QQ  contributions,
 $ U^{QQ}_{ij}(t)$, Eq. \eqref{U2QQ}, $ V^{QQ}_{ijk}(t)$, Eq. \eqref{V3tail}, $ U^{QQ}_{ijkl}(t)$, Eq. \eqref{U4tail},
 to the waveform. All propagators here are retarded.
}
\end{figure}

The physics behind this QQQ-coupling term (and its kins discussed below) is illustrated in Fig. \ref{fig:2}.

The LQ contribution $ U^{LQ}_{ij}(t)$ arises from an angular-momentum  
$\times$ quadrupole  $\times$ quadrupole coupling, and reads
\bea \label{U2LQ}
U_{ij}^{\rm LQ}(t)&=&\frac13 \frac{G}{c^5} \epsilon_{ab\langle i}I_{j\rangle a}^{(4)}L_b
\,,
\eea
where $L_b$ denotes the total angular momentum of the system.
The next term $ U^{WQ}_{ij}(t)$ arises from a WQQ cubic coupling. Here $W(t)= W_{i_1 \cdots i_\ell}$
 is the monopole component ($\ell=0$) of the time-gauge shift used outside the source to reduce 
 (at the first step of a MPM iteration) the exterior metric to the simple canonical form. It reads
\bea
U_{ij}^{\rm WQ}(t)&=& 4\frac{G}{c^5}  [W^{(2)}I_{ij}-W^{(1)}I_{ij}^{(1)}]^{(2)}\,.\qquad
\eea  
We have indicated  within parentheses  the memory contribution 
\be
U^{\rm memory}_{ij}(t)= -\frac27 \frac{G}{c^5} \int_{0}^{+\infty} d\tau
 I_{a\langle i}^{(3)}(t-\tau)I_{j\rangle a}^{(3)}(t-\tau)
 \ee 
in Eq. \eqref{U_ij_gen_all}   because it only contributes at order $O(G^3 \eta^5)$, which would correspond to the two-loop level.

Finally, one must take into account several radiation-reaction-level effects: (i) the time-domain source quadrupole moment $ I_{ij}(t)=Q_{ij}(t)+O(\eta^2)$
contains a 2.5PN ($\eta^5$) contribution (first computed in \cite{Blanchet:1996wx}); (ii) when evaluating the second derivative
of~\footnote{Here considered in the c.m. system, with $\mu =m_1 m_2/(m_1+m_2)$,
and $x^i(t) = x_1^i(t)-x_2^i(t)$ denoting the relative motion. The notation ${\rm STF}_{i_1 i_2 \cdots i_\ell}$
denotes the symmetric-trace-free projection of a Cartesian tensor.} $I^{\eta^0}_{ij}(t) =Q_{ij}(t)= \mu {\rm STF}_{ij} [x^ix^j]$
one must include the $G^2 \eta^5$ radiation-reaction contribution to the equations of motion
in harmonic coordinates (first computed in \cite{Damour:1981bh}); 
and finally, (iii) when evaluating the Fourier transform one must take
into account the radiation-reaction modification $\delta^{\rm rr} x^i(t) $ of the motion~\footnote{This last step can be by-passed if one works with $U_{ij}^{(1)}$.}.

Note that all the nonlinear  contributions to the radiative quadrupole (as well as the last radiation-reaction effect) 
are of order $G^2\eta^5$. Their Fourier transform can then be directly computed by inserting the Newtonian-level
straightline motion. By contrast the evaluation of the Fourier transform of $ U_{ij}^{\rm tail}$ requires using
 $\eta^2$-accurate motions. 

As an example of a radiation-reaction-level contribution, let us highlight the
  2.5PN ($G^2\eta^5$) contribution to the source quadrupole (with $n_{\rm orb}^i \equiv x^i/r$) \cite{Blanchet:1996wx}
  \be \label{Q2.5PN}
  I_{ij}^{\rm 2.5 PN}(t) = \frac{G^2 M^3}{c^5} \nu^2 \, {\rm STF}_{ij} \left[ -\frac{24}{7} \dot r \, n_{\rm orb}^i n_{\rm orb}^j + \frac{48}{7} v^i n_{\rm orb}^j\right]\,.
  \ee
There are similar time-dissymmetric contributions to all the higher multipole radiative moments $V_2, U_3, V_3, U_4$ entering the 
MPM waveform. In the present study we shall project the waveform on its even-in-$\phi$ part. As a
consequence, only $U_2$, $V_3$ and $U_4$ will be of interest to us (at the accuracy $\frac{G^2}{c^5}$).

In view of the fact that the higher multipolar contributions $V_3$ and $U_4$ enter the waveform with a prefactor
$\eta^2$, we need to include several other time-dissymmetric contributions (of fractional order $\eta^3$) to both 
 $V_3$ and $U_4$. The MPM formalism has given the following expressions for these contributions (we can again neglect
 memory effects in $U_4$):
 \beq
V_{ijk}= V_{ijk}^{\rm tail}+V_{ijk}^{\rm LQ + QQ}\,,
\eeq
with
\bea \label{V3tail}
V_{ijk}^{\rm tail}(t)&=& \frac{2G{\mathcal M}}{c^3}\int_0^\infty d\tau J_{ijk}^{(5)}(t-\tau)\left( \ln \left(\frac{\tau}{2 b_0} \right) + \frac{5}{3}\right)
\,,\nonumber\\
V_{ijk}^{\rm LQ + QQ}(t)&=&\frac{G}{c^3}\left(-2I_{\langle ij}^{(4)}L_{k\rangle}-\frac{1}{10}\epsilon_{ab\langle i}I_{j|a|}I^{(5)}_{k\rangle b}\right.\nonumber\\
& +& \left. \frac12 \epsilon_{ab\langle i}I^{(1)}_{j|a|}I^{(4)}_{k\rangle b}\right)\,,
\eea
while
\beq
U_{ijkl}=U_{ijkl}^{\rm tail} +U_{ijkl}^{\rm QQ} \;\; (+U_{ijkl}^{\rm memory})\,,
\eeq
with
\bea \label{U4tail}
U_{ijkl}^{\rm tail}(t)&=&\frac{2G{\mathcal M}}{c^3}\int_0^\infty I_{ijkl}^{(6)}(t-\tau) \left( \ln \left(\frac{\tau}{2b_0}\right) +\frac{59}{30}\right)
\,,\nonumber\\
U_{ijkl}^{\rm QQ}(t)&=&\frac{G}{c^3}\left(-\frac{21}{5}I_{\langle ij}I_{kl\rangle}^{(5)}-\frac{63}{5}I_{\langle ij}^{(1)}I_{kl\rangle}^{(4)}\right.\nonumber\\
&-&\left. \frac{102}{5}I_{\langle ij}^{(2)}I_{kl\rangle}^{(3)}
\right)\,.
\eea

\section{Brief reminder of the structure of the effective field theory (EFT) waveform expressions}

 Recent  quantum-based computations (which we shall label, for simplicity, as ``EFT" though they come in different flavours)  
 have derived the
classically-relevant part of the one-loop scattering amplitude with four scalars and one graviton \cite{Brandhuber:2023hhy,Herderschee:2023fxh,Georgoudis:2023lgf}. 
According to Refs. \cite{Brandhuber:2023hhy,Herderschee:2023fxh,Georgoudis:2023lgf} one can define~\footnote{In particular,
Ref. \cite{Brandhuber:2023hhy} asserts that their specific heavy-mass EFT formalism has the property of directly computing
the classically relevant part of the quantity 
$\langle p'_1p'_2 k^h | T | p_1p_2\rangle - i \langle p'_1p'_2|T^\dagger a^h(k) T |p_1p_2 \rangle$ yielding 
the Fourier-domain GW waveform.}
a classically-relevant part of the  (momentum-space) 5-point amplitude ${\mathcal M}(\eps, k, q_1, q_2) $, where $\eps$ is a null polarization vector, $k$ is
the null momentum of the emitted graviton, and $q_1=p_1-p'_1, q_2=p_2-p'_2$ are transferred momenta, such
that it is related (via the formalism of Ref. \cite{Kosower:2018adc}) to the (frequency-domain) classical waveform (say in the form
of the $W(\om,  \theta,\phi)$ in Eq. \eqref{Wom_th_phi}) via  (with {$\kappa \equiv \sqrt{32 \pi G}$, and $c=1$ in this section)
\be
{\mathcal M}^{\rm int}(k) = -\frac{ i \kappa}{2} W(\om,  \theta,\phi)\,.
\ee
Here $\om, \theta,\phi$ parametrize $k$, and  ${\mathcal M}^{\rm int}(k)$ denotes the following
 {\it integrated} version of the 5-point amplitude defined (in our signature) as
\be \label{Mint_first}
{\mathcal M}^{\rm int}(k) \equiv \int \mu(q_1,q_2,k) e^{- i (q_1\cdot b_1 + q_2 \cdot b_2)} {\mathcal M}(\eps, k, q_1, q_2)\,,
\ee
with a  measure $\mu(q_1,q_2,k)$ originally given (in the heavy-mass EFT \cite{Brandhuber:2023hhy},
and in dimension $D$) as
\bea
\mu_{{\bar p}}(q_1,q_2,k)&=& \frac{d^D q_1}{(2 \pi)^{D-1}} \frac{d^D q_2}{(2 \pi)^{D-1}} (2 \pi)^D \times \nonumber\\
&& \delta^D(q_1+q_2-k) \delta( 2 \,{\bar p}_1 \cdot q_1) \delta( 2 \,{\bar p}_2 \cdot q_2) \,,\nonumber\\
\eea
where the ``barred momenta variables" ${\bar p}_1, {\bar p}_2$ are defined in terms of the
incoming ($p_a$) and outgoing ($p'_a$) momenta variables entering the 5-point amplitude
 as \cite{Landshoff:1969yyn,Parra-Martinez:2020dzs}  
\be 
\label{pbar}
\bar p_a  \equiv \frac12 (p_a + p'_a)\,.
\ee
The associated momentum transfers $q_a \equiv p_a-p'_a$ are then
such that one has
\bea
p_1 &=& {\bar p}_1+ \frac12 q_1\,,\qquad p'_1 = {\bar p}_1- \frac12 q_1\, , \nn \\
p_2 &=& {\bar p}_2+ \frac12 q_2\,, \qquad  p'_2 = {\bar p}_2- \frac12 q_2\, .
\eea
The mass shell conditions $p_a^2=p'_a{}^2=-m_a^2$ then imply that $\bar p_a \cdot q_a=0$.

It is asserted in Refs. \cite{Brandhuber:2023hhy,Herderschee:2023fxh,Georgoudis:2023lgf} that, in the classical limit, one can neglect the formally $O(\hbar)$ difference between 
the ${\bar p}_a$'s and the incoming momenta $p_a$ and use as measure
\bea
\mu_{{ p}}(q_1,q_2,k)&=& \frac{d^D q_1}{(2 \pi)^{D-1}} \frac{d^D q_2}{(2 \pi)^{D-1}} (2 \pi)^D\times \nonumber\\ 
&& \delta^D(q_1+q_2-k) \delta( 2 \,{ p}_1 \cdot q_1) \delta( 2 \,{ p}_2 \cdot q_2) \,.\nonumber\\
\eea
However, we found (as detailed below) that, when comparing the EFT waveform with the MPM one, the two waveforms 
are closer if one defines the EFT waveform by using a
measure $\mu_{{\bar p}}(q_1,q_2,k)$ involving the {\it classical} barred momenta variables, together with corresponding definitions
of the impact parameters $b_1, b_2$. Here, by classical barred momenta we mean those defined by using Eq. \eqref{pbar} with the
{\it classical} outgoing ($p'_a$) momenta (differing from the incoming ones by the classical impulse $\Delta p_a=p'_a-p_a=O(G)$).

The integral defining ${\mathcal M}^{\rm int}(k)$ in Eq. \eqref{Mint_first} is Poincar\'e-covariant,  and depends 
(besides its indicated dependence on $k$) on: (i) the null polarization vector $\eps$ used to define a null spin-2 polarization tensor $\eps^{\mu\nu}=\eps^{\mu} \eps^{\nu}$; (ii) the  momenta ($p^\mu_a$ or ${\bar p}^\mu_a$) used in the integration measure $\mu$ and in the integrand, e.g. ${\mathcal M}(\eps, k, q_1, q_2, \bar p_1, \bar p_2)$; 
and (iii) the two impact parameters $b^\mu_1$ and $b^\mu_2$ (consistently taken in the 2-plane orthogonal to  $\bar p_1$, 
and $\bar p_2$). 

The MPM formalism has been mainly developed within the (incoming) center-of-mass (c.m.) system. In order to compare 
the MPM waveform $W^{\rm MPM}$ to the EFT one, $W^{\rm EFT}$ (defined from ${\mathcal M}^{\rm int}(\om, \theta,\phi)$), we need to chose  $b^\mu_1$ and $b^\mu_2$ 
so that they are mass centered. The c.m. system is predominantly defined by choosing as unit time vector
\be
e_0^\mu= \frac{p_1^\mu+p_2^\mu}{| p_1^\mu+p_2^\mu|}\,.
\ee
At the present
PM accuracy of the EFT waveform we do not need to worry about the recoil of the binary system, which is 
negligible both classically, when using  
$p'_1 + p'_2-p_1 -p_2=-P^{\rm rad}=O(G^3)$, and quantum mechanically, when
using  the quantum EFT constraint $q_1+q_2= p_1 + p_2-p'_1 -p'_2= k$,
with $k=O(\hbar)$ (for a single graviton). We can then equivalently define the c.m. system, when working
with classical barred momenta, by choosing as unit time-vector
\be
{\bar e}_0^\mu= \frac{{\bar p}_1^\mu+{\bar p}_2^\mu}{|{\bar  p}_1^\mu+{\bar p}_2^\mu|}= e_0^\mu+ O(G^3)\,.
\ee
We then choose the impact parameters  $b^\mu_1$ and $b^\mu_2$ to be orthogonal to $e_0^\mu$ (or ${\bar e}_0^\mu$)
and to satisfy
\be
\frac{E_1}{E } b_1^\mu + \frac{E_2}{E } b_2^\mu =0\,.
\ee
Here and in the following $E_1$, $E_2$ and $E\equiv E_1+E_2$ denote the (incoming) energies in the c.m. frame.

Defining the relative transferred momentum $q^\mu$ by
\bea
\label{q1_2_defs}
q_1^\mu &=& q^\mu + \frac{E_1}{E } k^\mu\,,\nonumber\\ 
q_2^\mu &=& -q^\mu + \frac{E_2}{E } k^\mu\,,
\eea
the exponential factor entering Eq. \eqref{Mint} reads
\be
 e^{- i (q_1\cdot b_1 + q_2 \cdot b_2)} = e^{- i q\cdot b_{12}} \,,
\ee
where $b_{12}^\mu = b_1^\mu-b_2^\mu$ is the relative vectorial impact parameter.

In order to define the function  $W^{\rm EFT}(\om, \theta,\phi)$ and compare it to its MPM analog 
$W^{\rm MPM}(\om, \theta,\phi)$ we need to fix the spatial direction (in the c.m. frame) of the various vectors defining
the GW waveform. A crucial issue is to define the angles of emission $\theta,\phi$ of the GW with respect
to the direction of the spatial momenta of the two particles. 
We found in this respect that there was a significant difference in using the integration measure
$\mu_{{ p}}$ (involving the incoming momenta), or an integration measure 
$\mu_{{\bar p}}$ (involving  the classical barred momenta).

The issue arises mainly when combining the tree and one-loop amplitudes and integrating over the transferred
momenta, say (using a measure $ \mu_{\bar p}$, and an integrand, involving the same variables ${\bar p}_a$)
\bea \label{Mintpbar}
&&{\mathcal M}^{\rm int}(k,\bar p_1, \bar p_2,b_{12}) =\int \mu_{\bar p}(q_1,q_2,k) e^{- i (q_1\cdot b_1 + q_2 \cdot b_2)}\nonumber\\ 
&&\qquad \times  \left[{\mathcal M}^{\rm tree}(\eps, k, q_1, q_2; \bar p_1, \bar p_2)\right.\nonumber\\
&&\qquad +\left. {\mathcal M}^{\rm one-loop}(\eps, k, q_1, q_2; \bar p_1, \bar p_2)\right]\,.
\eea
In the case indicated, the use of the $ \mu_{\bar p}$ measure implies (after expressing $q_1$ and $q_2$ in terms of the relative transferred
momentum $q$, Eq. \eqref{q1_2_defs}) that the spatial direction of $q$ in the c.m. frame is defined in terms of the 
spatial direction of the barred momenta. In addition,
any Poincar\'e-covariant amplitude is a function of various scalar products, involving (besides $\eps, k, q_1, q_2$)
some momenta variables. In Eq. \eqref{Mintpbar} we have used the $\bar p_a$ variables in all terms: the measure,
the tree amplitude, and the one-loop contribution. The relative impact parameter $b^\mu_1-b^\mu_2$ is then
consistently taken to be orthogonal to both $\bar p_1$ and $\bar p_2$.

In our EFT/MPM comparison we have explored two different choices for the momenta variables $\bar p_a$ used
in the computation of ${\mathcal M}^{\rm int}(k,\bar p_1, \bar p_2,b_{12})$, Eq. \eqref{Mintpbar}.
Being  aware that, within an in-in computation, e.g. using the framework of \cite{Kosower:2018adc}, the variables
$p'_a$ (entering the 5-point amplitude) are not the classical outgoing momenta but belong to the incoming wavepacket,
we started by using momenta variables  $\bar p_a= p_a +O(\hbar)$. However, when making this choice, 
  we found large differences between the EFT waveform and the MPM one, which started at the leading-Newtonian-order
  (see below). This is why we then focussed on computing the EFT waveform by using the classical values of 
 the  $\bar p_a$ momenta, thereby finding a much closer agreement between the two waveforms.

Given any choice of the momenta variables used to define the EFT waveform, the 
 emission angles $\theta,\phi$ of the GW are then defined as the  polar
angles with respect to the spatial frame $e_1$, $e_2$, $e_3$ anchored on the corresponding choice of momenta
variables. 

When using the incoming momenta (in the c.m. frame), the second axis $e_2$ is in the spatial direction of 
${p}_1$ (and is denoted $e_p$), while the first axis $e_1$ is in the direction of the incoming vectorial impact parameter $b^-_{12}$
(and is denoted $e_b$). The third axis $e_3$  is orthogonal to the plane of motion (and is always denoted $e_z$).

When using the classical barred momenta, we use the spatial frame  $e_1=e_x$, $e_2=e_y$, $e_=e_z $ used
 in Ref. \cite{Bini:2021gat} (see Eq. (3.49) there).
This frame~\footnote{which is also the frame used in eikonal studies, see  e.g. Ref. \cite{DiVecchia:2023frv}.} is defined such that  $e_y$ is  the bisector between the incoming and the outgoing spatial momentum
of the first particle (in the c.m. frame). This precisely corresponds to taking $e_y$ in the spatial direction of ${\bar p}_1$
rather than that of ${p}_1$. The axis $e_x$ is then taken in the direction orthogonal to ${\bar p}_1$ (in the direction
from 2 to 1). \typeout{to check}

Let us give details on the computation of the EFT waveform when using the latter ${\bar p}_a$-anchored frame.
Decomposing all vectors in the frame ${\bar e}_0$, $e_x$, $e_y$, $e_z$ we have
\bea
\label{frame_exey}
k &=& \om \left( {\bar e}_0 + {\bf n}(\theta,\phi) \right), \nn \\
{\bf n}(\theta,\phi) &=& \sin\theta \cos\phi \, e_x +\sin\theta\sin\phi \, e_y +\cos\theta \,e_z \,, \nn\\
q &=& q^0  {\bar e}_0 + q^x e_x +q^ye_y +q^z e_z\,, \nn \\
\eps &=& {\bf \mb}= \frac{1}{\sqrt{2}} \left(\D_\theta {\bf n}(\theta,\phi)-  \frac{i}{\sin \theta} \D_\phi {\bf n}(\theta,\phi)\right)\,,\nn\\
b_{12} &=& b^x e_x\,, \nn \\
{\bar p}_1 &=& {E}_1  {\bar e}_0 +  {\bar P}_{\rm c.m.} e_y\,,\nn \\
{\bar p}_2 &=& {E}_2  {\bar e}_0 -  {\bar P}_{\rm c.m.} e_y\,.
\eea
Here, $ {E}_a =\sqrt{\bar m_a^2 +{\bar P}_{\rm c.m.}^2 }= \sqrt{m_a^2 +{P}_{\rm c.m.}^2 }$, where ${\bar P}_{\rm c.m.} = {P}_{\rm c.m.}  \cos \frac12 \chi^{\rm class}$ differs from the standard
c.m. momentum ${P}_{\rm c.m.}$ only at order $\chi_{\rm class}^2= O(G^2)$. The latter difference will be negligible
in our comparison below. In other words, we can approximate 
${\bar m}_a^2 \equiv - {\bar p}_a^2=m_a^2 + P_{\rm c.m}^2 \sin^2 \frac{\chi}{2}= m_a^2+ O(G^2) $ by $m_a^2$.

The components $q^0$ and $q^y$ of $q$ are obtained (in view of the delta functions in $\mu_{\bar p}$) by solving
the two constraints $q_1 \cdot {\bar p}_1=0$ and $q_2 \cdot {\bar p}_2=0$. At the end the integration over 
$\mu_{{\bar p}}$ yields (in dimension $D=4$)
\bea \label{Mint}
{\mathcal M}^{\rm int}(\om, \theta,\phi) &=&\frac1{4 \bar m_1 \bar m_2 \sqrt{y^2-1}} \times \nonumber\\
&& \int \frac{dq^x}{2\pi} \frac{dq^z}{2\pi} e^{- i q^x b} {\mathcal M}(\eps, k, q_1, q_2)\nonumber\\
&=&  -\frac{ i \kappa}{2} W^{\rm EFT}(\om,  \theta,\phi).
\eea
Here $y \equiv  - {\bar p}_1 \cdot {\bar p}_2/({\bar m}_1 {\bar m}_2)$.

After having aligned our frame with the classical barred momenta, the dependence of the amplitudes on the
various $\bar p$-dependent scalars (such as $y$ versus $ \g \equiv - p_1 \cdot p_2/(m_1 m_2)$, 
$\bar m_a$ versus $m_a$, or the eikonal-type impact 
parameter $b^x$ entering Eqs. \eqref{frame_exey} versus the incoming impact parameter $b^-$, see Eq. \eqref{b_min_eq} below) is such that we can actually neglect the $O(G^2)$ fractional
differences in such scalars: Indeed, $y=\g + O(G^2)$,  $\bar m_a = m_a + O(G^2)$, $b^x=\frac{b^-}{\cos \frac{\chi_{\rm cons}}{2}}=b^-+ O(G^2)$,
etc. and
a fractional $O(G^2)$ correction in the tree amplitude is equivalent to a two-loop effect~\footnote{In the following we will generally denote $b^-=b^x+ O(G^2)$  as $b$.}. 
One should however keep in mind this issue when considering the amplitude at the two loop
level. [Even at the one loop level, there is an issue about the correct definition of
the zero-frequency contribution to the waveform, which would require a $G^3$ accuracy
rather than a $G^2$ one. As we shall only consider $\om \neq 0$, we shall not worry
here about this issue.]

\section{Combined multipolar, post-Minkowskian and post-Newtonian expansion of  $W^{\rm MPM}(\om,  \theta,\phi)$}

The comparison between the EFT waveform $W^{\rm EFT}(\om,  \theta,\phi)$ (defined by Eq. \eqref{Mint})
and $W^{\rm MPM}(\om,  \theta,\phi)$ (defined by Eq. \eqref{Wom_th_phi})
necessitates to compute some (Fourier-type)
integrals on both sides. We found both types of integrations to be somewhat demanding. As some the MPM building blocks
(notably the source multipole moments) have only been computed to some limited PN (rather than PM) accuracy, the
comparison between the waveforms must be done in a PN-expanded sense, in addition to the combined
multipolar post-Minkowskian expansion structure explicitly used in $W^{\rm MPM}(\om,  \theta,\phi)$,
and only implicitly present in $W^{\rm EFT}(\om,  \theta,\phi)$.

 Let us display the basic PN and PM structure
of the perturbatively computed waveform. It reads (in the classical normalization of $h_c =4 G  \eta^4 W$)
\bea
W(\om,  \theta,\phi)&=&\frac{ 1}{4 G \eta^4} \left[ G^1 A^{\rm lin}(\theta,\phi) \delta(\om) \right.\nonumber\\ 
&+& G^2 A^{\rm post-lin \, or\, tree}(\om,\theta,\phi)\nonumber\\
&+& G^3 A^{\rm post-post-lin \, or\, one-loop}(\om,\theta,\phi)\nonumber\\
&+&\left.  O(G^4) \right]\,,
\eea
Here, $G^1 A^{\rm lin}$ corresponds to the linear-gravity (1PM) value of $h_{\mu\nu}$, taken along  straight worldlines.
It is stationary, and yields a zero-frequency contribution to the frequency-domain waveform. This contribution is well-known
and is incorporated in the MPM formalism through the incoming and outgoing asymptotic values of the multipole moments
(or more precisely through the average between the incoming and outgoing asymptotic values of the waveform). In the following,
we only consider the strictly positive frequency axis $\om >0$, without considering the latter zero-frequency, static contribution.

The  PN-expansion of the postlinear (2PM) (or tree level)  contribution to $W=h_c/(4G\eta^4)$  starts  at the ``Newtonian order" ($\eta^0$).
Its leading PN contribution   is given  (in the time domain) by Einstein's quadrupole formula
\bea
W^{\rm post-lin \, or\, tree}(\om,  \theta,\phi)&=& U_2(\om) + O(\eta)\nonumber\\
&=& \int_{- \infty}^{+ \infty} dt e^{i \om t} U_2(t) +O(\eta)\,,\qquad
\eea
where 
\bea
\label{U2formal}
 U_2(t) &=&  \frac{1}{2!} \mb^{i} \mb^{j } I^{(2)}_{i j}(\om)+ O(G \eta^3)\nonumber\\
&=&  \frac{1}{2!} \mb^{i} \mb^{j } \frac{ d^2}{dt^2} \left[\mu x^i x^j+ O(\eta^2) \right] +O(G \eta^3)\,.\qquad
\eea
Here, we have indicated for concreteness the beginning of the MPM expansion of the (time-domain)
radiative quadrupole moment $U_{ij}(t)$. It starts with Einstein's LO quadrupole formula 
$U_{ij}^{LO}(t)= {\rm STF}_{ij} \frac{ d^2}{dt^2} \mu x^i x^j$, where we recall that we work in the c.m. system
and denote  $\mu =m_1 m_2/(m_1+m_2)$, $x^i(t) = x_1^i(t)-x_2^i(t) \equiv r n^i_{\rm orb}$.  
The 1PN-order correction, $+ O(\eta^2)$, 
to the quadrupole moment was first computed (within the MPM formalism) in Ref. \cite{Blanchet:1989ki}. 
The 1.5PN correction, $+O(G \eta^3)$, in Eq. \eqref{U2formal} denotes the (linear) tail contribution \cite{Blanchet:1992br,Blanchet:1997jj}. 
In spite of the absence of an explicit power of $G$ in the LO
radiative quadrupole moment $U_{ij}^{LO}(t)=  {\rm STF}_{ij} \frac{ d^2}{dt^2} \mu x^i x^j$, its frequency-domain version is (when
considering strictly positive frequencies) of order $G^1$. This is most simply seen by considering its time-derivative
$\frac{d}{dt} U_{ij}^{LO}(t)$, which necessarily involves one acceleration $a^i = \frac{ d^2}{dt^2} x^i= -GM \frac{n_{\rm orb}^i}{r^2} + O(\eta^2)$, and  reads, at leading order,
\be \label{dotQ2}
\frac{d}{dt} U_{ij}(t) =- 2 \frac{GM \mu}{r^2}\,   {\rm STF}_{ij} \left( 4 n^i_{\rm orb} v^j - 3 \dot r n^i_{\rm orb} n^j_{\rm orb} \right) + O(\eta^2)\,.
\ee
Here, the 1PN contribution $+ O(\eta^2)$ comes both from the 1PN contribution to the source quadrupole moment
and from the 1PN contribution to the acceleration $a^i = \frac{ d^2}{dt^2} x^i$. 
The frequency-domain version of Eq. \eqref{dotQ2} reads
\bea 
\label{dotQ2om}
- i \om  U_{ij}(\om)&=& - 2 GM \mu \, {\rm STF}_{ij} \int \frac{dt e^{i \om t}}{r^2} \left( 4 n_{\rm orb}^i v^j \right.\nonumber\\
&&\left. - 3 \dot r n_{\rm orb}^i n_{\rm orb}^j  + O(\eta^2)\right)\,.
\eea
To compute the $t$ integral on the  rhs we need to insert the explicit time-dependence of the (PN-expanded~\footnote{We recall
that the PN expansion incorporates a PM expansion in powers of $G$, e.g. 1PN$  \sim \eta^2 v^2 + G\eta^2 M/r$.} ) 
relative~\footnote{We systematically work in the c.m. frame, and express all multipole moments in terms of the
relative motion variables, $x^i(t)=x_1^i(t)-x_2^i(t), v^i(t)= d x^i(t)/dt$.} scattering motion. The latter is obtained
by  {\it perturbatively solving} the PN-expanded two-body equations of motion. The explicit form of the solution
of the equations of motion depends on the spatial frame in which they are expressed. 

There are two ways to compute the PN-expanded motion. A standard PM$+$PN-perturbation approach to hyperbolic motion would be to give initial data
at $t=- \infty$ (notably initial velocities) and to solve for perturbations of straight line motions, say
\bea \label{PMpert}
{\bf x}(t) &=& {\bf x}_0(t) + G {\bf x}_1(t) + G^2 {\bf x}_2(t) + \cdots\,, \nn\\
{\bf v}(t) &=& {\bf v}_0(t) + G {\bf v}_1(t) + G^2 {\bf v}_2(t) + \cdots \,,
\eea
where
\bea \label{b_min_eq}
 {\bf x}_0(t) &=& b^- e_b + v_0 t e_p\,, \nn \\
  {\bf v}_0(t) &=&  v_0  e_p\,.
\eea
Here, $ b^- $ denotes the incoming impact parameter (such that $L= P_{\rm c.m.} b^-$),
$ e_b$ denotes the unit vector  in the direction of the (incoming) vectorial impact parameter, while
$ e_p$ denotes the unit vector in the direction of the incoming velocity of the first particle in the c.m. frame.
As shown, e.g., in section IVA of Ref. \cite{Bini:2018ywr}, fixing some of the initial conditions at the moment
of closest approach of the two worldlines uniquely fixes the time scale entering the asymptotic logarithmic drift
of the worldlines in terms of the impact parameter.

A second way to compute the  PN-expanded motion is to use the {\it quasi-Keplerian} representation of binary motions
(introduced in \cite{DD85} 
at the 1PN level, and generalized to higher PN orders in \cite{Damour:1988mr,Cho:2018upo}). 
 This representation is
naturally expressed in the $e_x, e_y$ vectorial frame (which is naturally tuned to the classical ${\bar p_a}$ momenta, rather
than the incoming momenta $p_a$). It reads
\bea
{\bf x}(t) &=& x(t) e_x + y(t) e_y\nonumber\\
&=& r(t) \left( \cos \varphi(t) e_x + \sin \varphi(t) e_y \right)\,,
\eea
where the polar coordinates of the relative motion are obtained as functions of time by eliminating the
(hyperbolictype) ``eccentric anomaly" variable $v$ between equations of the type
\bea \label{QK}
{\bar n}\, t &=& e_t \sinh(v) - v + O(\eta^4)\,, \nn\\
r &=& {\bar a}_r \left( e_r \cosh(v) -1) \right) + O(\eta^4)\,, \nn \\
\varphi &=& 2 K \arctan\left(\sqrt{\frac{e_\varphi+1}{e_\varphi-1}}\tanh\left(\frac{v}{2}\right)\right)   + O(\eta^4)\,.
\eea
Here, the quasi-Keplerian quantities ${\bar n}, e_t, e_r, e_\varphi, K$ are (PN-expanded) functions of the
c.m. energy and angular momentum of the binary system (see Refs. \cite{DD85,Damour:1988mr,Cho:2018upo}).
The quasi-Keplerian representation Eq. \eqref{QK} incorporates (in the conservative case) a time symmetry around
$t=0$, corresponding to the closest approach between the two bodies. The asymptotic logarithmic drift of the two
worldlines is embodied in the $v$ parametrization involving hyperbolic functions.

Inserting the quasi-Keplerian representation Eq. \eqref{QK} in the Fourier-transform \eqref{dotQ2om} leads to
PN-expanded integrals of the type
\be
\int dv e^{i \frac{\om}{{\bar n}} \left(  e_t \sinh(v) - v  \right)} \left( F_0(v) + \eta^2  F_2(v) + \cdots \right)\,.
\ee
At the first few PN orders,  these integrals can be expressed (for a fixed, finite value of the eccentricity) in terms
of (modified) Bessel K functions of the type $K_{\nu_{QK}}(u_{QK})$ where the natural Quasi-Keplerian argument $u_{QK}$ is defined as
$u_{QK}= \frac{\om e_r}{\bar n}$ and where the Bessel order depends on the Bessel argument as $\nu_{QK}= i\frac{u_{QK}}{e_r}+n$, 
where $n=0$ or $1$ (see, e.g., \cite{Bini:2017wfr}). 

We are interested here in the PM re-expansion of these integrals, {\it i.e.} in their large impact parameter
expansion, corresponding to an expansion in inverse powers of the eccentricity $e_r$. When doing this secondary expansion
the order of the Bessel functions tends to $0$ or $1$, but we need to evaluate derivatives of $K_\nu(u)$ with respect to
the order $\nu$. The first $\nu$ derivative is easy to evaluate, but the second $\nu$ derivative introduces higher-level
special functions.  [This fact makes it challenging to compute frequency-domain multipoles at the 2PN level.]
Finally, using the PM (large-eccentricity) reexpansion of the PN-expanded Quasi-Keplerian representation 
we could explicitly evaluate all the  source multipole moments (up to rather high multipolar orders) both
at the Newtonian ($\eta^0$) fractional~\footnote{In PN theory it is traditional to qualify, say,
the leading PN approximation of the multipole moments as being of Newtonian level, because
they are given by Newtonian-looking expressions, e.g. $I_{i_1 \cdots i_\ell}= \mu {\rm STF} x^{i_1} \cdots x^ {i_\ell} + O(\eta^2)$.
Note, however, that their contribution to the waveform contains an extra factor 
$\eta^{\ell-2}$ (resp. $\eta^{\ell-1}$  for even (resp. odd) parity.}
accuracy, and at the 1PN  ($\eta^2$) fractional accuracy. 

Let us indicate here the structures of  the 1PN-accurate values
of the radiative multipole contributions $U_2$,  $V_3$ and $U_4$ computed by using the quasi-Keplerian
representation, and by referring all tensorial quantities (including $\epsilon$ and $k$, and therefore $\theta, \phi$)
to the frame  ${\bar e}_0$, $e_x$, $e_y$, $e_z$.
They are conveniently expressed in terms of
the symmetric mass ratio $\nu \equiv \frac{m_1 m_2}{M^2}$  (where $M\equiv m_1+m_2$)
and of the (dimensionless) frequency variable
\be
\label{u_def}
u\equiv \frac{\om \, b}{c \pinf} \; ; \; {\rm assuming} \; \om >0\,,
\ee
which enters as the argument of various (modified) Bessel functions. Here, we considered that $p_\infty$ is dimensionless (when restoring physical units), {\it i.e.}, $p_\infty=v_\infty/c$ with $v_\infty$ a velocity.

Displaying here for simplicity their structure when considering the emission in the equatorial plane ($\theta=\frac{\pi}{2}$)
we have (at the 1PN accuracy, {\it i.e.} neglecting time-symmetric 2PN contributions, as well as time-dissymmetric radiation-reaction 
and tail contributions)
\bea
U^{\rm equat}_2(\om, \phi)&=& U_2^{G {\eta^0}}(\om, \phi) +  U_2^{G {\eta^2}}(\om, \phi)\nonumber\\
&+&  U_2^{G^2 {\eta^0}}(\om, \phi) +  U_2^{G^2 {\eta^2}}(\om, \phi)\nonumber\\
&+& O(\eta^4) + O(G^2 \eta^3) + O(G^3)\,,
\eea
where
\bea
\label{U2_1PN_eqs}
U_2^{G {\eta^0}}&=&-\frac{GM^2\nu}{2}  \left( A_{20}^{G\eta^0}K_0(u)+A_{21}^{G\eta^0}K_1(u) \right)\frac{1}{p_\infty}\,, \nonumber\\
U_2^{G {\eta^2}}&=& -\frac{GM^2\nu}{2} \eta^2  
\left( A_{20}^{G\eta^2}K_0(u)+A_{21}^{G\eta^2}K_1(u) \right) p_\infty \,,\nonumber\\
U_2^{G^2 {\eta^0}}&=& \frac{\pi}{2} u \left(\frac{GM}{bp_\infty^2}\right)  U_2^{G \eta^0}\,,\nonumber\\  
U_2^{G^2 {\eta^2}}&=&  \frac{G^2M^3\pi \nu u}{b p_\infty}\eta^2 \left[ A_{20}^{G^2\eta^2}K_0(u)+A_{21}^{G^2\eta^2}K_1(u)\right.\nonumber\\
&+&\left. B_2^{G^2 {\eta^2}}e^{-u}  \right] \,.
\eea
Note the presence of a contribution $\propto e^{-u}$ at the $G^2 {\eta^2}$ level.

The coefficients of the Bessel $K_0(u)$ and $K_1(u)$ functions read
\bea
\label{U2_1PN_eqs_coeff}
A_{20}^{G\eta^0}&=&2i\sin(2\phi) u + \cos(2\phi) + 1 \,,\nonumber\\
A_{21}^{G\eta^0}&=& 2i \sin(2\phi) + 2 u\cos(2\phi) \,,\nonumber\\
A_{20}^{G\eta^2}&=& \left[\left(\frac{12}{7}\nu  - \frac{19}{21}\right) u^2 
+  \frac{4}{7}\nu  - \frac{13}{7}\right]\cos(2\phi) \nonumber\\
&+& \frac{6}{7}i\left(\nu + \frac{1}{18}\right) u\sin(2\phi) +  \frac{4}{7}\nu  - \frac{13}{7}\,,\nonumber\\
A_{21}^{G\eta^2}&=& \left(\frac{12}{7}\nu  - \frac{17}{42}\right) u\cos(2\phi)\nonumber\\ 
&+& \left[\left( \frac{12}{7}\nu  - \frac{19}{21} \right) u^2 +  \frac{8}{7}\nu   + \frac{23}{7}\right] i\sin(2\phi) \nonumber\\
&+& \left(\frac{8}{7}\nu  - \frac{59}{14}\right) u\,, \nn \\
A_{20}^{G^2\eta^2}&=&  -\frac{1}{168}(72 u^2\nu-38 u^2+45\nu-141)\cos(2\phi)\nonumber\\
&-&\frac{1}{84}i \sin(2\phi) (39\nu-62) u-\frac{15}{56}\nu+\frac{47}{56} \,,\nonumber\\
A_{21}^{G^2\eta^2}&=&  -\frac{1}{168}(114 u\nu-143 u)\cos(2\phi)\nonumber\\
&-&\frac{1}{84}i \sin(2\phi) (36 u^2\nu-19 u^2+45\nu+6)\nonumber\\
&-& \frac{2}{7} u\nu+\frac{59}{56} u \,,
\eea
while the coefficient of $e^{-u}$ is
\bea
B_2^{G^2 {\eta^2}}&=&  -\frac{3}{2u^2} [(u+u^2)\cos(2\phi)\nonumber\\
&+&\sin(2\phi)i(u^2+u+1)]\,. 
\eea

Similarly, we computed the 1PN-accurate values of $V_3(\om, \phi, \theta)$ and $U_4(\om, \phi,\theta)$.
Their structures in the equatorial plane read
\bea
\label{V3_struc}
V^{\rm equat}_3(\om, \phi)&=& V_3^{G {\eta^0}}(\om, \phi) +  V_3^{G {\eta^2}}(\om, \phi)\nonumber\\
&+&  V_3^{G^2 {\eta^0}}(\om, \phi) +  V_3^{G^2 {\eta^2}}(\om, \phi)\nonumber\\
&+& O(\eta^4) + O(G^2 \eta^3) + O(G^3)\,,
\eea
and
\bea
\label{U4_struc}
U^{\rm equat}_4(\om, \phi)&=& U_4^{G {\eta^0}}(\om, \phi) +  U_4^{G {\eta^2}}(\om, \phi)\nonumber\\
&+&  U_4^{G^2 {\eta^0}}(\om, \phi) +  U_4^{G^2 {\eta^2}}(\om, \phi)\nonumber\\ 
&+& O(\eta^4) + O(G^2 \eta^3)+ O(G^3)\,.
\eea
Let us only display here the Newtonian-level results, $O(G\eta^0)$ and   $O(G^2\eta^0)$, for both $V_3$ and $U_4$, 
relegating their 1PN contributions $O(G\eta^2)$ and   $O(G^2\eta^2)$ to Appendix \ref{1PN_corr_V3_U4}. Concerning $V_3$ we find
\bea
\label{V3_Newt}
V_3^{G {\eta^0}}&=& -p_\infty GM^2\nu u (1-3\nu) \left[A_{30}^{G\eta^0}K_0(u)+A_{31}^{G\eta^0}K_1(u)\right]\,,\nonumber\\
V_3^{G^2 {\eta^0}}&=&\frac{\pi}{2} u \left(\frac{GM}{bp_\infty^2}\right)  V_3^{G,\eta^0}\,,
\eea
where
\bea
A_{30}^{G\eta^0}&=& -\frac{1}{3} u\cos(2\phi)-\frac16 i \sin(2\phi)\,,\nonumber\\
A_{31}^{G\eta^0}&=&  -\frac13 i u \sin(2\phi) -\frac13 \cos(2\phi) \,.
\eea 
 
Concerning $U_4$ we find
\bea
\label{U4_Newt}
U_4^{G {\eta^0}}&=& -\frac{G\left(\nu - \frac13\right)M^2\nu p_\infty}{7}
\left[A_{40}^{G\eta^0}K_0(u)+A_{41}^{G\eta^0}K_1(u)\right]\,,\nonumber\\  
U_4^{G^2 {\eta^0}}&=& \frac{\pi}{2} u \left(\frac{GM}{bp_\infty^2}\right)  U_4^{G,\eta^0}\,,
\eea
where
\bea
A_{40}^{G\eta^0}&=& \frac{7}{2} i u \left(u^2+\frac{9}{4}\right) \sin (4 \phi )+\left(u^2+\frac{3}{2}\right) \cos (2 \phi )\nonumber\\
&+&\left(7 u^2+\frac{21}{8}\right) \cos (4 \phi )+\frac{9}{4} i u \sin (2 \phi
   )-\frac{9}{8}  \,, \nonumber\\
A_{41}^{G\eta^0}&=& \left(\frac{7 u^3}{2}+\frac{203 u}{16}\right) \cos (4 \phi )+\left(i u^2+3 i\right) \sin (2 \phi )\nonumber\\
&+& \left(\frac{35 i u^2}{4}+\frac{21 i}{2}\right) \sin (4 \phi )+\frac{11}{4} u \cos (2
   \phi )-\frac{15 u}{16}\,.\nonumber\\
\eea

\section{Computing the post-Newtonian expansion of  $W^{\rm EFT}(\om,  \theta,\phi)$, and its multipolar decomposition }

The EFT amplitude in $k,q_1,q_2$ space has the following structure
\bea
{\mathcal M}(k,q_1,q_2)&=&\kappa {\mathcal M}^{\rm lin}(k)
+\kappa^3 \bar m_1^2 \bar m_2^2 {\mathcal M}^{\rm tree}(k,q_1,q_2)\nonumber\\
&+& \kappa^5   \left(\bar m_1^3 \bar m_2^2 {\mathcal M}_1^{\rm one\,loop}(k,q_1,q_2)+1\leftrightarrow 2\right)\,,\nonumber\\
\eea
with  
\bea
{\mathcal M}^{\rm lin}(k)&=&-\frac{i}{2}\epsilon_\mu \epsilon_\nu \frac12 \sum_a m_a (u_a^\mu u_a^\nu 2\pi\delta (-k\cdot u_a)e^{-ik\cdot b_a}\nonumber\\ 
&&+u'_a{}^\mu u'_a{}^\nu 2\pi\delta (-k\cdot u'_a)e^{-ik\cdot b_a'} )\,,
\eea
and
\bea
{\mathcal M}^{\rm tree}&=&\frac{i}{4 q_1^2 q_2^2 w_1 w_2}\left[ F_3 (-4 F_3 w_1 w_2 + 2 F_3 (q_1^2 + q_2^2) y \right.\nonumber\\
&-& 4 F_2 w_1 y + 4 F_1 w_2 y)  \nonumber\\
&+&  \left(-\frac12  F_2 F_3 (q_1^2 + q_2^2) w_1 + F_2^2 w_1^2 \right.\nonumber\\ 
&+&\left.\left.
    \frac12  F_1 F_3 (q_1^2 + q_2^2) w_2 + F_1^2 w_2^2\right) (1 - 2 y^2)\right] \,.
\eea
Here, $q_1^2\equiv q_1\cdot q_1$, $q_2^2\equiv q_2\cdot q_2$ and (denoting $\bar u_a \equiv \bar p_a/\bar m_a$)
\bea
w_1 &=& -k\cdot \bar u_1\,,\qquad  
w_2 = -k\cdot \bar u_2\,,\qquad y = -\bar u_1\cdot \bar u_2\,,
\nonumber\\  
F_1&=& F^k_{\alpha\beta}q_1^\alpha \bar u_1^\beta\,,\quad
F_2= F^k_{\alpha\beta}q_2^\alpha \bar u_2^\beta\,,\quad  
F_3= F^k_{\alpha\beta}\bar u_1^\alpha \bar u_2^\beta\,,\nonumber\\
\eea
with the notation
\beq
F^k_{\alpha\beta}=[k\wedge \epsilon]_{\alpha\beta}=k_\alpha \epsilon_\beta-k_\beta \epsilon_\alpha\,.
\eeq

The $\bar m_1^3 \bar m_2^2$ piece of the one loop amplitude can be split into \lq\lq divergent" and \lq\lq finite" 
parts~\footnote{We call here ``divergent" part of ${\mathcal M}(k,q_1,q_2)$  the ($\mu_{\rm IR}$-regularized) part 
of the one-loop amplitude that is proportional to the tree amplitude, obtained after dropping the infinite $\frac1{\eps}$ contribution.}, namely
\begin{widetext}
\bea
\label{Eqs_amp_generali}
{\mathcal M}^{\rm one\, loop}_{\rm div}|_{\bar m_1^3 \bar m_2^2}&=&\left(\frac{1}{64}\frac{3y-2y^3}{(y^2-1)^{3/2}}w_1
+\frac{w_1}{32}+\frac{iw_1}{16\pi}\ln \left(\frac{w_1}{\mu_{\rm IR}}\right)\right) {\mathcal M}^{\rm tree} \,,\nonumber\\
{\mathcal M}^{\rm one\, loop}_{\rm fin}|_{\bar m_1^3 \bar m_2^2}&=& \frac{R}{\pi}  + c_1{\mathcal I}_1+c_2{\mathcal I}_2  
+ \frac{l_q}{\pi}\ln\left(\frac{q_1^2}{q_2^2}\right)  +\frac{l_{w_2}}{\pi}\ln \left(\frac{w_2^2}{w_1^2}\right) +\frac{l_y}{\pi \sqrt{y^2-1}}\ln (y+\sqrt{y^2-1})\,,
\eea
\end{widetext}
with
\bea
{\mathcal I}_1&=&   - \frac{i}{32\sqrt{q_2^2+w_1^2}} \nonumber\\ 
&+& \frac{1}{16\pi \sqrt{q_2^2 +w_1^2}}\ln\left( \frac{w_1+\sqrt{q_2^2+w_1^2}}{\sqrt{q_2^2}}\right)\,,\nonumber\\
{\mathcal I}_2&=& -\frac{i}{32 \sqrt{q_1^2}}\,.
\eea

Here the various coefficients $R$, $c_1$, $c_2$, $l_q$, $l_{\omega_2}$, $l_y$, are rational functions of the eight variables
$(w_1, w_2, q_1^2, q_2^2,  y, F_1,F_2, F_3)$, defined above.

The conditions
\beq
q_1\cdot \bar p_1=0\,,\qquad q_2\cdot \bar p_2=0\,,
\eeq
determine the two components $q^0$ and $q^y$.
Neglecting henceforth the $O(G^2)$ differences between $\bar m_a$ and $m_a$, and $y$ and $\gamma$, but keeping the information about the frame $e_x-e_y$, the solution for $q^0$ and $q^y$ reads
\bea
q^0 &=& \omega\frac{ (m_1 m_2\sqrt{\gamma^2 - 1}n^y - m_1^2 + m_2^2)}{(2\gamma m_1 m_2 + m_1^2 + m_2^2)}\,,\nonumber\\
q^y&=& \omega \frac{(\gamma ^2+1) m_1 m_2 + \gamma (m_1^2 +  m_2^2) }{\sqrt{\gamma^2 - 1}(2\gamma m_1 m_2 + m_1^2 + m_2^2)}\,. 
\eea
Integrating out $q^0$ and $q^y$ yields the following Jacobian
\beq
\int dq^0 dq^y \delta (2q_1\cdot \bar p_1)\delta (2q_2\cdot \bar p_2) =\frac{1}{4 m_1m_2 p_\infty}\,.
\eeq

We recall the relations
\bea
E_1 &=& \frac{m_1}{E} (\gamma m_2 + m_1)\,,\nonumber\\  
E_2 &=& \frac{m_2}{E} (\gamma m_1 + m_2)\,, 
\eea
with $E_1+E_2=E=Mh$ and
\bea
P_{\rm cm} = \frac{m_1 m_2}{E}\sqrt{\gamma^2 - 1} \,,\quad h=\sqrt{1+2\nu(\gamma-1)}\,.
\eea

The PN expansion ({\it i.e.} the expansion in powers of $\pinf$) of ${\mathcal M}^{\rm tree}$  leads to  
sum of terms with rational numerators and denominators of the type
\beq
\label{polynomtree}
\frac{P_m(q_x,q_z; \omega)}{\om^\vareps \left[ p_\infty^2 [(q^x)^2 + (q^z)^2]  + \omega^2 \right]^{n} } \,.
\eeq
Here $P_m(q_x,q_z; \omega)$ denotes a homogeneous polynomial of  degree $m$ in the variables $q_x,q_z; \omega $;
the exponent $\vareps$ of $\om$ in the denominator is either 1 or 0; the exponent $n$ is a positive integer; and the
overall homogeneity degree $m - \vareps- 2n$ is equal to $-2$.

It is then convenient to rescale the $q_x,q_z$ variables as
\beq
q_x=\Omega Q_x\,,\qquad q_z=\Omega Q_z\,,
\eeq
where
\beq
\Omega\equiv\frac{\omega}{p_\infty}\,.
\eeq
The PN-expanded tree amplitude is then expressed as a sum of rational
terms having integer powers of $D_0 \equiv 1+Q_x^2+Q_z^2$ as denominators, with overall homogeneity $\Omega^{-2}$.
Suppressing the dependence of the unintegrated ${\mathcal M}^{\rm tree}$ on $ Q_x,Q_z$, the angles, and the masses,
we have the $\Omega$ and $\pinf$ structure
\bea \label{Mtreeexp}
 \kappa^{-1}{\mathcal M}^{\rm tree}(\Omega, p_\infty, Q_x,Q_z)&=&\frac{G}{\Omega^2}\left(1+p_\infty+p_\infty^2\right.\nonumber\\
&+&\left. \pinf^3+\ldots\right)\,.
\eea

Using the known  integral (valid even when the exponent $n$ of $D_0 \equiv 1+Q_x^2+Q_z^2$ is not an integer)
\beq \label{intD0n}
\int \frac{dQ_xdQ_z}{(2\pi)^2}\frac{e^{-iuQ_x}}{D_0^n}=\frac{2^{-n} u^{n-1}}{\pi  \Gamma(n)}K_{n-1}(u)
\eeq
(and relations between Bessel functions)
the integrated versions of
${\mathcal M}^{\rm tree}(k,q_1,q_2)$ 
(and thereby ${\mathcal M}^{\rm one\, loop}_{\rm div} \propto {\mathcal M}^{\rm tree}(k,q_1,q_2)$)
can be explicitly expressed as a sum of Bessel functions $K_{0}(u)$ and $ K_{1}(u)$,
with argument
\be \label{udef}
u \equiv  \Omega b  \equiv\frac{\om b}{\pinf} \,,
\ee
 and with polynomial coefficients in $u$. Note that the dimensionless variable $u$ introduced here
 is the same as the one used above  in the PN expansion of the MPM waveform, see Eq. \eqref{u_def}.

By contrast, the contribution to the waveform coming from
${\mathcal M}^{\rm one\, loop}_{\rm fin}$ is more challenging to evaluate.
Indeed, the coefficients entering ${\mathcal M}^{\rm one\, loop}_{\rm fin}$ involve large expressions containing spurious poles. Moreover, the location of these poles is PN-dependent. In performing the PN expansion
of  ${\mathcal M}^{\rm one\, loop}_{\rm fin}$, checking along the way that all the spurious poles cancel, one ends up
with a structure of a type similar  to the one in Eq. \eqref{polynomtree}, but with the following significant
differences: (i) the exponent $n$ of  $ p_\infty^2 [(q^x)^2 + (q^z)^2]  + \omega^2 $ in the
denominator is either an integer or a half integer; and (ii) the overall homogeneity  degree $m - \vareps- 2n$ is equal to $-1$.
For the same reason as above, all the terms in  ${\mathcal M}^{\rm one\, loop}_{\rm fin}$ involving integer powers
of   $ p_\infty^2 [(q^x)^2 + (q^z)^2]  + \omega^2 $ ({\it i.e.} integer powers of $D_0 \equiv 1+Q_x^2+Q_z^2$ after
the $\Omega$ rescaling) will yield, after integration on $Q_x, Q_z$, polynomial combinations of Bessel functions $K_{0}(u)$ and $ K_{1}(u)$
with argument \eqref{udef}. By contrast, the contributions involving denominators with  half-integer values of $n$ yield,
after  integration on $Q_x, Q_z$, terms of the form $P_m(u) e^{-u}$ (which correspond, in the general-$n$ formula
Eq. \eqref{intD0n} to Bessel K functions of half-integer orders).

Suppressing the dependence of the unintegrated ${\mathcal M}^{\rm one\, loop}_{\rm fin}$ on $ Q_x,Q_z$, the angles, and the masses,
we then find that it has the following $\Omega$ and $\pinf$ structure
\bea
\label{eq:fin_G2overOmega}
\kappa^{-1}{\mathcal M}^{\rm one-loop, fin}(\Omega, p_\infty, Q_x,Q_z)&=&\frac{G^2}{\Omega}\left(1+ p_\infty+ \pinf^2\right.\nonumber\\
&+&\left.\pinf^3 +\ldots\right)\,.
\eea
For instance, the unintegrated $O(\bar m_1^3 \bar m_2^2)$  ${\mathcal M}_{\rm fin}^{\rm one\, loop}$, at order $p_\infty^0$ and on the equatorial plane,  reads
\begin{widetext}
\bea
&&{\mathcal M}_1^{\rm fin,\,p_\infty^0}(Q_x,Q_z; \phi, \theta=\frac{\pi}{2})= \frac{(-\frac{3}{512} Q_x\sin(2\phi)i + \frac{3}{512}\cos(2\phi)i + \frac{3}{256}\cos(\phi) Q_z)}{\Omega D_0^{1/2}}\nonumber\\
&&+\frac{(-\frac{3}{1024} Q_x^2\cos(2\phi)i - \frac{3}{512} Q_x\sin(2\phi)i + \frac{9}{1024} Q_x^2i - \frac{3}{256} Q_z Q_x\sin(\phi) + \frac{3}{1024}\cos(2\phi)i + \frac{3}{256}\cos(\phi) Q_z + \frac{9}{1024}i)}{\Omega D_0^{3/2}}\,.
\eea
\end{widetext}

Compared to the tree-level amplitude Eq. \eqref{Mtreeexp}, the finite part of the one-loop amplitude differs by an
overall factor of order (putting back the total mass $M=m_1+m_2$ for dimensional reasons)
\be
\frac{{\mathcal M}^{\rm one-loop,fin}}{{\mathcal M}^{\rm tree}} \simeq G M \Omega=\frac{G M \omega}{\pinf}=
\frac{GM}{b}\,u=O \left( \frac1{c^2}\right)\,.
\ee
As indicated, the finite part of the one-loop amplitude differs from the tree amplitude by an overall 1PN-order factor
(of the $\frac{GM}{c^2 r}$ type).

On the other hand, taking into account the value of the factor ${\mathcal F}^{\rm one\, loop}\equiv \frac{{\mathcal M}^{\rm one-loop,div}}{{\mathcal M}^{\rm tree}} $, namely
\bea
\label{one_loop_div}
{\mathcal F}^{\rm one\, loop}(\om, p_\infty; \theta,\phi)
&=& \frac{\pi}{2} G E\omega \frac{3\gamma-2\gamma^3}{p_\infty^3}\nonumber\\
&+& 2G E\omega \left(\frac{\pi}{2}+i\ln \left(\frac{|\omega|}{\mu_{\rm IR}}\right)\right) \nonumber\\
&+& 2i  G \left[ m_1\omega_1\ln \left(\frac{|\omega_1|}{|\omega|}\right)\right.\nonumber\\
&+&\left. m_2\omega_2\ln \left(\frac{|\omega_2|}{|\omega|}\right)\right]\,,
\eea
which starts at the Newtonian order 
\be
{\mathcal F}^{\rm one\, loop} \sim \frac{G M \om}{\pinf^3}=\frac{GM \Omega}{\pinf^2}
=O \left( \frac1{c^0}\right)\,,
\ee
 (as further discussed below), the divergent part of the amplitude, 
\bea
{\mathcal M}^{\rm div}(\Omega, p_\infty,   Q_x,Q_z)&=&{\mathcal F}^{\rm one\,loop}{\mathcal M}^{\rm tree}\,,\qquad
\eea
has the following $\Om$, $\pinf$ structure
\beq
\kappa^{-1}{\mathcal M}^{\rm div}(\Omega, p_\infty, Q_x,Q_z)=\frac{G^2}{\Omega}\left( \frac{1}{p_\infty^2}+\frac{1}{p_\infty}+1+\ldots\right) .
\eeq

For simplicity, when evaluating the one-loop amplitude, we mainly focussed on the even-in-$\phi$ sector of the waveform ({\it i.e.}, $\propto \sum_m e^{i2 m\phi} $) by averaging ${\mathcal M}(\theta,\phi)$ and ${\mathcal M}(\theta,\phi+\pi)$.
In addition, motivated by the appearance of principal-value (PV) propagators for the massive particles in the heavy-mass EFT
formalism, we decided to focus
 on few terms having as fractional PN order (with respect to the leading-order quadrupole formula): Newtonian ($\eta^0$), 1PN ($\eta^2$), 1.5PN ($\eta^3$) and 2.5PN ($\eta^5$). Indeed, the 2.5PN order is the order where many radiation-reaction effects, and
 retarded-radiation effects, enter the MPM formalism, and we wanted to test whether the EFT results would contain these
 effects in spite of their use of time-symmetric PV propagators for the massive particles.
In the PN expansion of the finite part of the one-loop amplitude,
\bea 
\kappa^{-1}{\mathcal M}^{\rm fin}(\Omega, p_\infty, Q_x,Q_z)&=&\frac{G^2}{\Omega}\left(\pinf^0+p^1_\infty \right.\nonumber\\ 
&+&\left. p^2_\infty+ \pinf^3+\cdots \right)\,,\qquad 
\eea
the 2.5PN level is contained in the  $p_\infty^3$ term, while the   LO term 
$ \sim \pinf^0$  corresponds to the 1PN level.

The restriction to  the even-in-$\phi$ sector  has the property of projecting out all the even parity multipoles $U_l$ for $l$ odd  and all the odd parity multipoles $V_l$ for $l$ even.
Together with the restriction to the maximum 2.5PN fractional order this means that the multipolar, and PN, content of the resulting EFT waveforms is
\beq
W_{\rm EFT}\sim U_2^{\le \rm 2.5PN}+\eta^2 V_3^{\le \rm 1.5PN}+\eta^2 U_4^{\le \rm 1.5PN}\,,
\eeq
{\it i.e.}, when comparing to its MPM analog, we are reaching up to the 2.5PN fractional accuracy of the radiative quadrupole,
($\eta^5$ beyond the Newtonian quadrupole) and up to the  1.5PN fractional accuracy of the radiative $3^-$ and $4^+$ 
multipoles.

\section{Comparison between EFT and MPM  at the Newtonian order}

\subsection{Structure of the EFT waveform at the Newtonian order $(G+G^2) \eta^0$}

In the one-loop accurate result  
\be
{\mathcal M}=\left(1+{\mathcal F}^{\rm one\, loop}\right){\mathcal M}^{\rm tree}+{\mathcal M}^{\rm fin}\,,
\ee
the factor multiplying ${\mathcal M}^{\rm tree}$ has the following Newtonian, {\it i.e.}, nonrelativistic limit ($E\to M$, $\gamma\to 1$, $p_\infty\to 0$)
\be 
1+{\mathcal F}^{\rm one\, loop}=1+\frac{\pi}{2}\frac{GM\omega}{p_\infty^3}+O(\eta^2)\,.
\eeq
When restoring ordinary physical units (and notably the velocity of light) keeping $p_\infty\frac{v_\infty}{c}$ dimensionless, we have
\beq
1+{\mathcal F}^{\rm one\, loop}=1+\frac{\pi}{2} \frac{GM \omega}{v_\infty^3} +O\left(\frac{v_\infty^2}{c^2}\right)\,,
\eeq
where the dimensionless factor $\frac{\pi}{2} \frac{GM \omega}{v_\infty^3}$ contains no $c$'s in the nonrelativistic limit, $\frac{1}{c}\to 0$.
This shows that the tree-plus-one-loop divergent part of the amplitude admits a nonrelativistic limit in which the divergent part corrects the tree amplitude by the (real) Newtonian factor $1+\frac{\pi}{2} \frac{GM \omega}{v_\infty^3} $. In addition, Eq. \eqref{eq:fin_G2overOmega} shows that the finite contribution is smaller by a factor $p_\infty^2=v_\infty^2/c^2$. As a consequence, the inclusion of the one loop contribution in the non relativistic limit of the amplitude is predicted by the EFT result to be
\beq
\left(W^{\rm tree}+W^{\rm one\, loop}\right)^{\rm nonrelativistic}=\left(1+\frac{\pi}{2} \frac{GM \omega}{v_\infty^3} \right){\mathcal M}^{\rm tree}\,.
\eeq
More precisely, as the angular dependence of ${\mathcal F}^{\rm one\, loop}$ is only contained in the term (see Eq. \eqref{one_loop_div})
\beq
2i  G \left[ m_1\omega_1\ln \left(\frac{|\omega_1|}{|\omega|}\right)+ m_2\omega_2\ln \left(\frac{|\omega_2|}{|\omega|}\right)\right]\,,
\eeq
which is $ p_\infty^3=v_\infty^3/c^3$ smaller than the leading order contribution, this implies that we have the following Newtonian-level amplification factor
\beq
\left(W^{\rm tree}_\ell+W^{\rm one\, loop}_\ell\right) \overset{\frac{1}{c}\to 0}{=} \left(1+\frac{\pi}{2} \frac{GM \omega}{v_\infty^3} +O\left(\frac{1}{c^2}\right)\right)W_\ell^{\rm tree}
\eeq
separately for  {\it each multipole}. 

An important point is that the Newtonian-level EFT predictions just discussed hold independently of the choice of
momentum variables ${\bar p}_a$ used to define the EFT waveform. In other words, if one uses (as recommended in
Refs. \cite{Brandhuber:2023hhy,Herderschee:2023fxh,Georgoudis:2023lgf}) ${\bar p}_a=p_a+O(\hbar)$, the one-loop
results predict that the Newtonian-level waveform is described by (recalling that $u = \frac{\om b}{\pinf}$)
\be \label{W1loopNewt}
W^{(G +G^2)\eta^0}_{\rm EFT}(\om, \theta, \phi) = \left( 1+\frac{\pi}{2} u\, \frac{GM}{b p_\infty^2} \right) W_{\rm tree}^{G \eta^0}(\om, \theta, \phi),
\ee
where an easy computation shows that the leading Newtonian-order tree-level waveform is given by the same Newtonian-level
result obtained above from computing the Fourier transform of the MPM amplitude, namely (in the equatorial plane)
\bea \label{WtreeNewt}
W_{\rm tree}^{G \eta^0}(\om, \frac{\pi}{2}, \phi) &=& -\frac{GM^2\nu}{2 \pinf}\times \nonumber\\
&& \left[  (2i u\sin(2\phi)  + \cos(2\phi) + 1) K_0(u)\right.\nonumber\\ 
&+&\left.   (2i \sin(2\phi) + 2 u\cos(2\phi))  K_1(u) \right]\,.
\eea
In this prediction, one should (when using ${\bar p}_a=p_a+O(\hbar)$) refer all vectorial quantities, $\epsilon$ and $k$ (and therefore $\theta, \phi$), 
to the incoming frame $e_b, e_p$. 

By contrast, when using the classical values of the barred momenta ${\bar p}_a=p_a+O(G \eta^0)$ to define the
EFT waveform, the one-loop results predict that the Newtonian-level waveform is described by the {\it same} function
$W^{(G +G^2)\eta^0}_{\rm EFT}(\om, \theta, \phi)$ defined by Eqs. \eqref{W1loopNewt}, \eqref{WtreeNewt}, but
with a different definition (and physical meaning) of the angles $\theta, \phi$, and a correspondingly physically different polarization vector
\bea \label{barm}
\epsilon_1=\bar m_1 &=&\frac{1}{\sqrt{2}}  (\cos(\theta)\cos(\phi) + i\sin(\phi) ) \,, \nonumber\\
\epsilon_2=\bar m_2 &=& \frac{1}{\sqrt{2}}(\cos(\theta)\sin(\phi) - i\cos(\phi))\,,  \nonumber\\
\epsilon_z=\bar m_z &=& -\frac{1}{\sqrt{2}} \sin(\theta)\,,
\eea
where the axes $e_1, e_2$ are either $e_b, e_p$, or $e_x, e_y$.
Because of this, the EFT prediction, Eqs. \eqref{W1loopNewt}, \eqref{WtreeNewt}, can only be true
with respect to (at most) one of the spatial frames $e_b, e_p$, or $e_x, e_y$.

A first (physically rooted) way to see that the Newtonian-level EFT prediction defined by Eqs. \eqref{W1loopNewt}, \eqref{WtreeNewt}
is only true when using the classical barred momenta frame, $e_x, e_y$,
rather than the incoming momenta frame, $e_b, e_p$, is to note that, 
at the Newtonian quadrupolar level, the waveform emitted during an hyperbolic encounter exhibits
as preferred direction of emission (in the c.m. frame) the direction $ e_y$ defined in terms of classical barred momenta (which is
the mean direction between the incoming and the outgoing momenta), rather than the direction of the
incoming momenta, $e_p$. E.g., the radiated linear momentum at the fractional Newtonian accuracy 
(given by Eq. (G7) in \cite{Bini:2021gat}) is directed along $e_y$, and not along $e_p$. 

 A second way to see that the the EFT prediction, Eqs. \eqref{W1loopNewt}, \eqref{WtreeNewt}, is only
 true with respect to the  $e_x, e_y$ frame is to compare Eqs. \eqref{W1loopNewt}, \eqref{WtreeNewt}
 with the the quasi-Keplerian computation of the MPM waveform
 anchored to  $e_x, e_y$ already detailed in Section IV above.
 Indeed, when comparing the quasi-Keplerian
MPM results for the $G \eta^0$ and  $G^2 \eta^0$ contributions to $U_2$, given in Eqs.  \eqref{U2_1PN_eqs},
one precisely recognizes the agreement with Eqs. \eqref{W1loopNewt}, \eqref{WtreeNewt}. 
More generally, the Newtonian-level real amplification factor
\beq
1+\frac{\pi}{2}\frac{GM}{bp_\infty^2} \, u \,.
\eeq
predicted by the EFT results is seen to be present in the higher multipolarity contributions
$V_3$ and $U_4$ , given in Eqs. \eqref{V3_Newt} and  \eqref{U4_Newt}. 
As part of our EFT-MPM comparison we have explicitly checked that this property holds for the Newtonian-level $G+G^2$ MPM waveform up to $l=7$. 

As a third way of confirming the conclusion that, in order to have a close agreement between the EFT waveform
and the MPM one one needs to anchor the former  to classical ${\bar p}_a$ variables, we performed
a direct computation of the  MPM waveform anchored to the $e_b, e_p$ frame, as we explain now.

\subsection{Structure of the MPM waveform at  $O(G+G^2) \eta^0$ in the incoming $e_b, e_p$ frame}

Directly computing the (Newtonian-order)  MPM waveform in the incoming $e_b, e_p$ frame means using
the standard PM$+$PN-perturbation approach to hyperbolic motion indicated in Eqs. \eqref{PMpert}}, \eqref{b_min_eq}
above. Working at the Newtonian order,  $O(G+G^2) \eta^0$, one perturbative solution ${\bf x}(t)= x(t) e_b + y(t) e_p$
of the hyperbolic motion reads
\bea \label{solxyperteta0}
x(t)&=&b +\frac{GM}{b v_0^2} \left(- v_0 t - \sqrt{b^2 + v_0^2 t^2}   \right), \nn \\
y(t)&=& v_0 t + \frac{GM}{ v_0^2} \arcsinh \frac{v_0 t}{b}\,.
\eea
Inserting this solution in the Newtonian-level expression of $U_{ij}^{(3)}(t)$ (which is of order $O(G)$),  computing its
Fourier transform, dividing by $\frac{d}{dt}=- i \om$, and contracting with ${\bar m}^i {\bar m}^j$, with Eqs. \eqref{barm}
(all indices being referred to the $e_b , e_p, e_z$ frame) yields a result for $W^{(G +G^2)\eta^0}_{\rm MPM}(\om, \frac{\pi}{2}, \phi)$ which is {\it different} from  the $(e_b , e_p, e_z)$-frame EFT prediction, Eqs. \eqref{W1loopNewt}, \eqref{WtreeNewt}.
The difference between the two $(e_b , e_p, e_z)$-frame waveforms 
 can be written in the following form (in the equatorial plane)
\bea \label{dWeta0}
W^{(G +G^2)\eta^0}_{ e_b, e_p, \rm EFT}(\om, \phi)- W^{(G +G^2)\eta^0}_{ e_b, e_p, \rm MPM}(\om, \phi)=\nn \\
 i \om \delta t \, W^{G \eta^0}(\om, \phi) + \delta \phi \frac{\D}{\D \phi} W^{G \eta^0}(\om, \phi)  \,,
\eea
with
\bea \label{dtdphi}
\delta t &=& \frac{GM}{v_0^3},\nn \\
\delta \phi &=&\frac{GM}{b v_0^2} = \frac{\chi}{2} +O(\eta^2) ,
\eea
where $\chi= \frac{2 GM}{b v_0^2} $ denotes the (Newtonian-order) scattering angle.

The first contribution $ i \om \delta t \, W^{G \eta^0}(\om, \phi)$ to the difference \eqref{dWeta0} is physically unimportant
because it can be absorbed by shifting the origin of time in the Newtonian perturbative  solution \eqref{solxyperteta0}.
In fact, it is entirely due to the fact that, contrary to the quasi-Keplerian solution where the origin of time was chosen
to correspond to the closest approach between the two bodies, the perturbative solution \eqref{solxyperteta0} 
is such that the two bodies are closest at the time $t_{\rm closest}= \frac{GM}{v_0^3}= \delta t$. If  we shift the time
variable used in the solution \eqref{solxyperteta0} to $t^{\rm new}= t- \frac{GM}{v_0^3}$, we find that the term 
$ i \om \delta t \, W^{G \eta^0}(\om, \phi)$ disappears from the difference \eqref{dWeta0}. 

By contrast, the second contribution $+ \delta \phi \frac{\D}{\D \phi} W^{G \eta^0}(\om, \phi)$ to the difference \eqref{dWeta0}
cannot be similarly gauged-away. Indeed, though it does simply correspond to the fact that rotating the frame $ e_b, e_p$ 
by half the scattering angle turns it into the  $ e_x, e_y$ frame, the presence of this term in Eq. \eqref{dWeta0} does say
that the EFT waveform computed from the incoming momenta differs from the MPM waveform computed from the
incoming momenta at the Newtonian level. This is our third way of seeing that, in order to be as close
as possible to the MPM waveform, one needs to define the EFT waveform
by consistently using classical barred momenta (in the measure and in the integrand). 
Note that there are  no rotational ambiguities in the definition of the MPM waveform. Actually, if we always fix the time-origin
in the MPM solution to the instant of closest approach, Eq. \eqref{dWeta0}, with $\delta t=0$, applies to the difference between
the MPM waveform computed in the $ e_x, e_y$ frame and to the MPM waveform computed in the $ e_b, e_p$ frame,
and says that the two waveforms are the same physical object, referred to two different frames.

\section{Agreement between MPM and EFT at the fractional 1PN order}

\begin{widetext}

In view of the results of the previous section, we shall henceforth always compute the EFT waveform by consistently
using the classical barred momenta in its definition \eqref{Mintpbar}, so that it agrees with the MPM one at the
Newtonian order. Starting from this leading-order agreement, we now explore higher PN levels, i.e. higher powers of $\eta$.

Let us now consider the 1PN level contributions, at order $G^2 \eta^2$, to the waveform $W(\om, \phi)$.
For the MPM multipoles contributing at this order we have already given  above  $U_2^{G^2 \eta^2}$, Eqs. \eqref{U2_1PN_eqs} and  \eqref{U2_1PN_eqs_coeff},  whereas 
we have listed in Appendix  \ref{1PN_corr_V3_U4} the analogous contributions for
 $V_3^{G^2 \eta^2}$, Eq. \eqref{V3_1PN}, and $U_4^{G^2 \eta^2}$, Eq. \eqref{U4_1PN}, 
used to form the associated \lq\lq MPM amplitude".

From the amplitude side at the same level, instead, we have obtained the following {\it $Q$-integrated} amplitudes~\footnote{The integration over $Q_x, Q_z$ adds the overall Jacobian factor $\frac{\Omega^2}{4 m_1 m_2 \pinf}$ which modifies
the $\pinf$ scaling.}
\bea
{\mathcal M}^{p_\infty^{-1}\rm EFT}_{\rm finite \,, int}&=& \frac{3\nu e^{-u}\kappa G^2\pi}{4 u b p_\infty}
(-(u^2 + u + 1)\sin(2\phi) + i(u + 1) u\cos(2\phi)) \,,\nonumber\\
{\mathcal M}^{p_\infty^{-1}\rm APM}_{\rm 1loop-div\,, int}&=& \frac{\kappa \pi G^2 u \nu}{192 b p_\infty}\left[
\left(  ((48 u^2 + 18 )\nu - 16 u^2 - 6 )i\cos(4\phi)+ ((-24 u^3 - 54 u)\nu + 8 u^3 + 18 u)\sin(4\phi)\right.\right.\nonumber\\
&+&\left. ((96 u^2 + 36)\nu - 40 u^2 - 84)i\cos(2\phi)  + (-84\nu u + 84 u)\sin(2\phi) + 18i\nu - 78i\right)K_0(u)\nonumber\\
&+&
\left( ((24 u^3 + 87 u)\nu - 8u^3 - 29u)i\cos(4\phi) + ((-60 u^2 - 72)\nu + 20 u^2 + 24)\sin(4\phi) \right.\nonumber\\
&+& \left.\left. (132\nu u - 104 u)i\cos(2\phi)  + ((-96 u^2 - 72)\nu + 40 u^2)\sin(2\phi)
+ 21i\nu u - 99i u\right)K_1(u)
\right]\,,
\eea
plus a term of the same order coming from the tree-level
\bea
{\mathcal M}^{p_\infty^{-1}\rm EFT}_{\rm tree\,, int}&=&\frac{\kappa G\nu}{4p_\infty}
\left[(-2 u\sin(2\phi) + i(\cos(2\phi)+1))K_0(u)  +  (2iu\cos(2\phi)  - 2\sin(2\phi))K_1(u)\right]\,.
\eea
As explained above, the presence of $e^{-u}$, or  $K_{0,1}(u)$ functions comes from having integrated
over $Q_x, Q_z$, terms containing half-integer, or integer, powers of $D_0=1+Q_x^2+Q_z^2$.

Transforming the EFT interated amplitudes in terms of their corresponding waveform, using
\beq
{\mathcal M}^{\rm EFT}=-\frac{i\kappa}{2}\left(U_2^{p_\infty^{-1}\rm 1PN}+\eta^2 (U_4^{p_\infty^{-1}\rm N}+V_3^{p_\infty^{-1}\rm N})\right)\,,
\eeq
one finds that the difference MPM-EFT turns out to vanish at this (time-symmetric) 1PN level.

\end{widetext}

\section{Disagreement between MPM and EFT at the radiation-reaction order}

By contrast with the successful result of the MPM-EFT comparison at the  {\it time-symmetric} fractional Newtonian and 1PN orders,
we show in the present section that there are significant differences in the two waveforms when considering {\it time
dissymmetric} contributions.

As examples of tail contributions to the frequency-domain MPM waveform \cite{Blanchet:1993ec}, let us display the tail contributions
to $U_2(\om)$ and $U_4(\om)$:
\bea
\label{eq:U2_tail}
U_{2}^{\rm tail}(\om)&=&  \frac{G^2M^3}{c^3b^2}\nu  p_\infty  u^2 \left[A_{2}^{K_0} K_0(u)
+A_{2}^{K_1} K_1(u)  \right]\times \nonumber\\
&&\left(-\frac{\pi}{2|\omega|} -\frac{i}{\omega} \left(\ln (2 b_0 |\omega|) + \g_E -\frac{11}{12}\right)\right)\,,
\eea
with
\begin{widetext}
\bea
A_{2}^{K_0}&=& 1+\eta ^2p_\infty^2 \left(\frac{15 \nu }{14}-\frac{13}{7}\right) 
+\left[1+\eta ^2 p_\infty^2 \left(\frac{12}{7} \left(\nu -\frac{19}{36}\right)
   u^2+\frac{15 \nu }{14}-\frac{13}{7}\right)\right]\cos(2\phi)\nonumber\\
&&
+\left[2   +\frac{13}{7}  \eta ^2 \left(\nu +\frac{1}{39}\right) p_\infty^2  \right]\, i u\sin(2\phi)
\,,\nonumber\\
A_{2}^{K_1}&=&\frac{8}{7} \eta ^2p_\infty^2 u \left(\nu -\frac{59}{16}\right) 
+\left[2  +\frac{19}{7} \eta ^2 \left(\nu -\frac{17}{114}\right) p_\infty^2  \right]u \cos(2\phi)\nonumber\\
&&
+\left[2  +\eta ^2 p_\infty^2 \left(\frac{12}{7}\left(\nu -\frac{19}{36}\right)
   u^2+\frac{15 \nu }{7}+\frac{23}{7}\right)\right]i\sin(2\phi)\,.
\eea

\bea
\label{eq:U4_tail}
U_{4}^{\rm tail}(\om)&=& -(1-3\nu) \frac{G^2M^2 }{4c^5b^2}\nu p_\infty^3 u^2  \left[A_{4}^{K_0} K_0(u)
+A_{4}^{K_1} K_1(u)  \right]\left(-\frac{\pi}{2|\omega|} -\frac{i}{\omega}\left(\ln ( 2 b_0|\omega|)+\gamma_E-\frac{59}{30} \right)\right)\,,
\eea
with
\bea
A_{4}^{K_0}&=&  \left(\frac{8 u^2}{21} + \frac{4}{7}\right)\cos(2\phi) + \left(\frac{8 u^2}{3} + 1\right)\cos(4\phi) + \frac{6}{7}iu \sin(2\phi)    +  \left(\frac{4}{3}i u^2  + 3i\right)u\sin(4\phi)- \frac{3}{7}
\,,\nonumber\\
A_{4}^{K_1}&=&  \frac{22}{21} u\cos(2\phi)  + \left(\frac{4}{3} u^2 + \frac{29}{6} \right)u\cos(4\phi) + \left(\frac{8}{21}i u^2  + \frac{8}{7}i\right)\sin(2\phi) + \left(\frac{10}{3}i u^2 + 4i\right)\sin(4\phi) - \frac{5 u}{14}\,.
\eea

As examples of terms generated by cubically nonlinear radiative effects in the MPM formalism, let us display
\bea
U_2^{\rm QQ}(\om)&=& \frac{\mu^2 M G^2 p_\infty^2 u}{c^5b}\left[\left((-\frac{8iu^2}{21} + \frac{11i}{21})\cos(2\phi) -  \frac{5u}{14}\sin(2\phi)  + \frac{11i}{21}\right)K_0(u)\right.\nonumber\\
&+&\left.\left((\frac{8 u^2}{21} - \frac{4}{21})\sin(2\phi) + \frac{iu}{6}\cos(2\phi)  + \frac{31 i u}{14}\right)K_1(u)  \right]\,,
\eea
and
\bea
U_4^{\rm QQ}(\om)&=&\frac{G^2M \mu^2  p_\infty^2u }{80c^5b}\left\{\left[13i\cos(4\phi) + (-28u^2 - 33)u\sin(4\phi) +\frac{52i}{7}\cos(2\phi) -  \frac{66}{7}u\sin(2\phi)  - \frac{39i}{7} \right]K_0(u)\right.\nonumber\\
&+&\left.
\left[\left(28iu^2 + \frac{87i}{2}\right)u\cos(4\phi)+ (-14u^2-40)\sin(4\phi) +\frac{66i}{7}u\cos(2\phi) -  \frac{80}{7}\sin(2\phi) -\frac{45i}{14}u\right]K_1(u)\right\}\,.
\eea

\end{widetext}

On the side of the EFT amplitude (and its associated waveform) the main technical difficulty consisted in extending the
PN expansion of the finite part of the one-loop to the $\pinf^3$ level beyond the leading-order  $\pinf^0$ term.
We recall that, as seen above, the leading order contribution to ${\mathcal M}^{\rm fin}$ is at the 1PN
order, namely it contains a factor $\pinf^2$ with respect to the leading order in ${\mathcal M}^{\rm div}$
(which is itself of the same ``Newtonian" order as the tree amplitude). 

Having extracted the even-in-$\phi$ sector of both the MPM and the EFT amplitudes leads to a
 simplified multipolar structure for both amplitudes: $W^{\phi-\rm even}=U_2+V_3+U_4$.
Knowing this simplified structure of ${\mathcal M}^{\rm fin}$  allowed
us to compute it as a function of $Q_x, Q_z$, the masses, and the angles. After adding the divergent part,
and Fourier transforming  over $Q_x, Q_z$  we could compute the $G^2(\eta^3 + \eta^5)$,
 time-asymmetric, part of $ W^{\rm EFT}= \frac{2 i}{\kappa} {\mathcal M}^{\rm one-loop}$.
 
 Computing the resulting difference
 \bea \label{diffW}
 \delta^{G^2(\eta^3 + \eta^5)} W(b,\om, \theta, \phi) &\equiv& W^{\rm EFT , G^2(\eta^3 + \eta^5)}(b,\om, \theta, \phi)\nonumber\\
&-&W^{\rm MPM , G^2(\eta^3 + \eta^5)}(b,\om, \theta, \phi)\qquad
 \eea
 we first found that one could get agreement at the $O(G^2\eta^3)$ level ({\it i.e.} the level of the leading-order quadrupolar tail, Eq. \eqref{U2tail})
simply by relating  the (arbitrary) MPM  time scale $2 \, b_0$ (used in the MPM formalism to scale the logarithm
 kernel $\ln(\frac{c \tau}{2 b_0})$ entering the tail integrals; see Eqs. \eqref{eq:U2_tail} and \eqref{eq:U4_tail}) and the (arbitrary) frequency scale $\mu_{\rm IR}$
 used in the EFT approach to scale the frequency-domain tail logarithms present in the divergent part of the one-loop
 amplitude, see Eq. \eqref{Eqs_amp_generali}$_1$ through
 \be \label{bmu}
 \ln (2 b_0 \mu_{\rm IR}) + \g_E= \frac12\,.
 \ee
However, after using the relation \eqref{bmu}, we found that the difference $ \delta^{G^2(\eta^3 + \eta^5)} W(b,\om, \theta, \phi)$
{\it did not vanish} at order $O(G^2\eta^5)$. Actually, we found that the remaining difference at order $G^2\eta^5$ contained
multipolar contributions of degree, and parity, $\ell^{\pm}= 2^+, 3^-, 4^+$. More precisely, we found that 
$ \delta^{G^2\eta^5} W(b,\om, \theta, \phi)$ was the frequency version of a time-domain waveform difference, namely
\be \label{deltaW1}
 \delta^{G^2\eta^5} W(b,\om, \theta, \phi) = \int dt e^{ i \om t} \left( \delta U_2(t)+ \eta^2 \delta V_3(t)+ \eta^2\delta U_4(t)  \right),
\ee
with corresponding STF tensors (denoting $l^k=\frac{L^k}{\mu}=\epsilon_{krs}x^rv^s$ as well as using $v^{ij}=v^iv^j$, $n_{\rm orb}^{ij}=n^i_{\rm orb}n^j_{\rm orb}$, etc.)  
\bea \label{deltaW2}
\delta U_{ij}(t) &=&\frac{G^2 M^3}{c^5}  \frac1{r^2} \left[ a_1 \dot r v^{ij}+(a_2v^2+a_3\dot r^2)v^in_{\rm orb}^j\right.\nonumber\\
&+&\left. \dot r(a_4v^2+a_5\dot r^2)n_{\rm orb}^{ij} \right]^{\rm STF}\,,\nonumber\\
\delta V_{ijk}(t) &=& \frac{G^2 M^3}{c^3}  \frac1{r^3} \left[\left(b_1v^{ij}+b_2\dot r v^i n_{\rm orb}^j\right.\right.\nonumber\\
&+&\left.\left. (b_3 v^2 +b_4\dot r^2)n_{\rm orb}^{ij}\right)l^k  \right]^{\rm STF}\,,\nonumber\\
\delta U_{ijkl}(t) &=&\frac{G^2 M^3}{c^3}  \frac1{r^2}  \left[ c_1 v^{ijk}n_{\rm orb}^l+ c_2\dot rv^{ij}n_{\rm orb}^{kl}\right.\nonumber\\
&+&  (c_3v^2+c_4\dot r^2)v^i n_{\rm orb}^{jkl}\nonumber\\
&+&\left. (c_5v^2+c_6\dot r^2)\dot r n_{\rm orb}^{ijkl}
 \right]^{\rm STF}\,.
\eea
These expressions for  $\delta U_{ij}(t)$, $\delta V_{ijk}(t)$ and $\delta U_{ijkl}(t)$ represent the most general time-odd
$O(G^2\eta^5)$ expressions for multipole differences having the correct dimension. They are generally parametrized
by fifteen dimensionless coefficients $a_i$, $b_i$, $c_i$, a priori depending only on the symmetric mass ratio
 $\nu = m_1 m_2/M^2$

The dimensionless coefficients $a_i$, $b_i$, $c_i$ respectively entering $\delta U_2(t)$, $\delta V_3(t)$ and $\delta U_4(t)$
were found to be uniquely\footnote{Because of the time-shift freedom present in the MPM solution, say 
$\delta t = c_t \frac{GM}{c^5} v_{\infty}^2$, there is a small
gauge-ambiguity in the determination of the two coefficients $a_2$ and $a_4$ of the form $\delta a_2=- 8 c_t \nu$,
$ \delta a_4=+ 6 c_t \nu$.} determined as the following quadratic polynomials in $\nu$:
\bea
a_1(\nu)&=&  24\nu+\frac{40}{21}\nu^2\,,\nonumber\\
a_2(\nu)&=&-8\nu-\frac{208}{21}\nu^2\,,\nonumber\\
a_3(\nu)&=&-24\nu-\frac{24}{7}\nu^2\,,\nonumber\\
a_4(\nu)&=&12\nu+\frac{64}{7}\nu^2\,,\nonumber\\
a_5(\nu)&=&0\,,
\eea
\bea
b_1(\nu)&=&8\nu^2\,,\nonumber\\
b_2(\nu)&=&-12\nu^2\,,\nonumber\\
b_3(\nu)&=&0\,,\nonumber\\
b_4(\nu)&=&0\,,
\eea
\bea
c_1(\nu)&=&-96 \nu^2\,,\nonumber\\
c_2(\nu)&=&72\nu^2\,,\nonumber\\
c_3(\nu)&=&0\,,\nonumber\\
c_4(\nu)&=&0\,,\nonumber\\
c_5(\nu)&=&0\,,\nonumber\\
c_6(\nu)&=&0\,.
\eea
 The fact that all these extra multipole contributions to the waveform are of the form $  a \nu + b \nu^2$ is
 compatible with the expected analytical structure of such $\frac{1}{c^5}$ contributions.
 They should contain as least a factor $\nu$ like all multipoles in the c.m. frame.

 The presence of  differences of order $\nu^2$ in $\delta U_2(t)$ would not have surprised us.
 Indeed, radiation-reaction-related effects contribute terms $O(\nu^2)$ in $U_2$,  as
 exemplified on Eq. \eqref{Q2.5PN}.
However, the presence of  differences of order $\nu^1$  in $U_2$ 
came as a surprise. Indeed, such a difference would be already visible  in the
probe limit of a test mass scattering off a massive object. 

The absence of  linear-in-$\nu$ contributions in the $b_i(\nu)$ and  $c_i(\nu)$ coefficients
parametrizing $\delta V_3(t)$ and $\delta U_4(t)$ are equivalent to confirming the corresponding (harmonic)
numbers $h_{3^-}=\frac{5}{3}$ and  $h_{4^+}=\frac{59}{30}$ respectively entering the corresponding
linear tail terms $V_{ijk}^{\rm tail}$, Eq. \eqref{V3tail}, and $U_{ijkl}^{\rm tail}$, Eq. \eqref{U4tail}.
Let us recall that the presence of such $\ell$-dependent rational numbers in the tail 
has been known for a long time \cite{Blanchet:1989ki,Blanchet:1995fr}  and re-derived within other formalisms \cite{Poisson:1994yf,Almeida:2021jyt}.

The precise values of these numbers, together
with $h_{2^+}=\frac{11}{12}$,   Eq. \eqref{U2tail}, correspond to the use
of an harmonic coordinate system. But, when changing the
coordinate system they are shifted by a common rational number \cite{Blanchet:2013haa}
so that the
differences $h_{3^-}- h_{2^+}=\frac{3}{4}$ and $h_{4^+}- h_{2^+}=\frac{21}{20}$ (which are
the only relevant numbers after having related $\mu_{\rm IR}$ 
to $2 b_0$, Eq. \eqref{bmu}) are coordinate independent.

\section{Soft expansions of the MPM and EFT waveforms} \label{secsoft}

Here, we consider the asymptotic behavior, at large (positive or negative) retarded times,
of the waveform 
\be
W_{\mu \nu}(t, {\bf n}) = \frac{c^4}{4 G} \lim_{R\to \infty} R\, h_{\mu \nu}(t, R, {\bf n}),
\ee
emitted by classical scattering solutions of Einstein's equations. Refs. \cite{Saha:2019tub,Sahoo:2020ryf,Sahoo:2021ctw} showed (see also references therein) that $W_{\mu \nu}(t, {\bf n}) $ behaves, when $t \to \pm \infty$, as
\bea \label{tsoft}
W_{\mu \nu}(t, {\bf n})   \overset{t \to \pm \infty}{= } &&
{\sf A}^{\pm}_{\mu \nu}({\bf n}) + \frac1t  {\sf B}^{\pm}_{\mu \nu}({\bf n}) 
+ \frac{\ln |t|}{t^2}  {\sf C}^{\pm}_{\mu \nu}({\bf n})
\nn \\
&+& \frac{1}{t^2} {\sf D}^{\pm}_{\mu \nu}({\bf n})+ o\left( \frac{1}{t^2}\right),
\eea
and  derived universal, exact expressions for the first three coefficients of the expansion \eqref{tsoft} in terms of asymptotic dynamical data (in a harmonic gauge where ${\sf A}^{-}_{\mu \nu}({\bf n})=0$). The difference 
$ \left[ {\sf A}_{\mu \nu}\right]={\sf A}^{+}_{\mu \nu}-{\sf A}^{-}_{\mu \nu}$ is the leading-order
memory. It is a function of the incoming and outgoing momenta (including the momenta of the finite frequency
gravitational radiation emitted during the scattering process). The notation 
$A_{\mu \nu}, B_{\mu \nu}, C_{\mu \nu}, F_{\mu \nu},  G_{\mu \nu}$
used in Refs. \cite{Saha:2019tub,Sahoo:2020ryf,Sahoo:2021ctw} is related to ours, after contraction with 
$ \epsilon^\mu \epsilon^\nu$ ($A_{\epsilon \epsilon}\equiv \epsilon^\mu \epsilon^\nu A_{\mu \nu}$, etc.), via: 
\bea
 A_{\epsilon \epsilon}&=& 2 G ( {\sf A}^+_{\epsilon \epsilon}-  {\sf A}^-_{\epsilon \epsilon})\,,\nonumber\\
 B_{\epsilon \epsilon}&=& 2 G  {\sf B}^+_{\epsilon \epsilon}\,,\nonumber\\ 
  C_{\epsilon \epsilon}&=& 2 G  {\sf B}^-_{\epsilon \epsilon}\,,\nonumber \\
   F_{\epsilon \epsilon}&=& 2 G  {\sf C}^+_{\epsilon \epsilon}\,,\nonumber\\
  G_{\epsilon \epsilon}&=& 2 G  {\sf C}^-_{\epsilon \epsilon}\,.
  \eea

  The time-domain asymptotic behavior given in Eq. \eqref{tsoft} implies a corresponding 
low-frequency asymptotic behavior of the non analytic part of the frequency-domain waveform 
$W_{\mu \nu}(\omega, {\bf n}) = \int_{- \infty}^{+ \infty} dt e^{i \omega t} W_{\mu \nu}(t, {\bf n}) $.
Namely, the ``soft expansion" (as $\omega \to 0^+$) of $W(\omega, {\bf n}) = \epsilon^\mu \epsilon^\nu W_{\mu \nu}(\omega, {\bf n}) $ reads (modulo regular contributions as $\omega \to 0^+$)
\be\label{soft0}
 W(\om) \overset{\om \to 0^+}{=} \frac{{\mathcal A}_0}{\om} + {\mathcal B}_0 \ln  \om + {\mathcal C}_0 \om (\ln \om)^2  
  + {\mathcal D}_0 \om \ln \om + \cdots
 \ee
 where, for instance,
 \bea
 {\mathcal A}_0 &=& i ( {\sf A}^+_{\epsilon \epsilon}-  {\sf A}^-_{\epsilon \epsilon}), \nn \\
  {\mathcal B}_0 &=& -( {\sf B}^+_{\epsilon \epsilon}-  {\sf B}^-_{\epsilon \epsilon}), \nn \\
   {\mathcal C}_0 &=& \frac{i}{2} ( {\sf C}^+_{\epsilon \epsilon}-  {\sf C}^-_{\epsilon \epsilon}).
   \eea
   
  In order to define the coefficient
  of the sub-sub-sub-leading soft term $+ {\mathcal D}_0 \om \ln \om$ in presence of the  previous
  one, $ {\mathcal C}_0 \om (\ln \om)^2 $, one must clarify the dependence of the waveform $ W(\om) $ on
  various frequency scales.

  The MPM waveform (defined from the quasi-Keplerian solution) jointly depends on the characteristic frequency scale of the two-body scattering,
  say $\om_c \equiv \frac{\pinf}{b}$ (where $b$ is the impact parameter), and on
  the arbitrary time scale $b_0$ entering
  the definition of the retarded time  (and consequently entering all the tail integrals). As a consequence of the latter
  fact, $ W^{\rm MPM}(\om) $ contains a factor $\exp \left ( i \frac{2 G {\cal M \om}}{c^3} \ln b_0 \right)$. 
We can then write $ W(\om) $ in
  the product form $ W^{\rm MPM}(\om)  = \exp \left ( i \frac{2 G {\cal M \om}}{c^3} \ln \frac{ b_0 c \pinf}{b}\right) \widehat W( \frac{\om b}{c \pinf})$, where $\widehat W$ depends only on the  adimensionalized frequency $\hat \om \equiv \frac{\om b}{c \pinf}$ (which
 coincides with the dimensionless argument $u$ used above). 
We can then rewrite the soft expansion, Eq. \eqref{soft0}, of the MPM waveform in the form
 \bea
\label{our_soft}
&& W^{\rm MPM}(\om) \overset{\om \to 0}{=} \exp \left ( i \frac{2 G {\cal M \om}}{c^3} \ln \frac{ b_0 c \pinf}{b}\right) \nn \\&&\qquad\times\left[\frac{{\mathcal A}}{\om} + {\mathcal B} \ln \hat \om + {\mathcal C} \om (  \ln \hat \om)^2  + {\mathcal D} \om \ln \hat \om + \cdots \right]\,,\qquad
 \eea
 where the frequencies entering the logarithms have been adimensionalized: $\hat \om \equiv \frac{\om b}{c \pinf}$.
 
 We have computed the coefficients $\cal A, \cal B, \cal C, \cal D$ entering the so-defined low-frequency expansion of the MPM
 waveform at the accuracy at which we worked (namely, $(G+G^2 ) \times(\eta^0 + \eta^2 +\eta^3 + \eta^5 )$, and even-in-$\phi$).  
 We then compared our PN-expanded results for $\cal A, \cal B, \cal C$ to the PN-expansion of the 
 corresponding universal classical soft expansion of \cite{Sahoo:2020ryf,Sahoo:2021ctw},
 which is discussed in some detail in Appendix \ref{app:B}. 
 [We discuss below the next  soft-expansion coefficient $\cal D$ for which no 
 universal, exact expression in terms of asymptotic dynamical data has been derived.] 
 As the MPM waveform $W^{\rm MPM}(\om)$ is explicitly derived by solving the
 classical Einstein equations, we expect that it will  satisfy 
 (at the perturbative accuracy at which it is computed) 
 the classical soft theorems that derive from 
 the large retarded time behavior of scattering solutions of Einstein's equations. 
 We indeed found perfect agreement with the classical soft theorems
 within our $(G+G^2 ) \times(\eta^0 + \eta^2 +\eta^3 + \eta^5 )$ accuracy. 

Concerning the leading-order memory terms,  we got (working in the frame of Eqs. \eqref{frame_exey}, and restricting for simplicity henceforth to the equatorial plane $\theta=\pi/2$)  
 \bea
{ \cal A}^{\rm MPM}&=&\frac{GM^2}{b} \nu i\left[-   \sin(2\phi)\right.\nonumber\\
&-&\left.  \frac12 \left( (2\nu + 3) \sin(2\phi)   +(3\nu - 1) \sin(4\phi) \right)\eta^2 p_\infty^2   \right]\nonumber\\
&-& \frac{3}{2} \frac{G^2M^3}{b^2}\eta^2 i\nu \pi\sin(2\phi)\,,
 \eea
 which agrees, within the accuracy indicated above~\footnote{Note the vanishing of the $O(\eta^3)$ and $O(\eta^5)$ contributions.}, with the corresponding standard 
 memory term, $\propto \frac1{2G}\eps^\mu \eps^\nu  A_{\mu\nu}$ in Eq. (1.6) of  \cite{Sahoo:2021ctw}, see Appendix \ref{app:B}. [Note the factor $ \propto \frac1G$ when translating from the soft expansion of $h_{\mu\nu}(\om)$ to that of $W(\om)$.]
 Similarly, we found  
  \bea
 { \cal B}^{\rm MPM}&=&  \frac{GM^2}{p_\infty}\nu \left[\frac{(\cos(2\phi) + 1)}{2} \right.\nonumber\\
&+&  \eta^2  p_\infty^2 \left(\frac12 (\nu - 2)  \cos(2\phi)  + \frac18 (3\nu - 1)\cos(4\phi)\right.\nonumber\\
  &+&\left.\left.  \frac{\nu - 7}{8}\right)\right]\nonumber\\
&+& \frac{2 G^2M^3\eta^3\nu}{b} \sin(2\phi)\nonumber\\
&+& \frac{ G^2M^3\eta^5\nu}{b}p_\infty^2[3(1+\nu)\sin(2\phi)\nonumber\\
&+&(3\nu-1)\sin (4\phi)] 
 \,,
 \eea
 which agrees with the corresponding result ${\cal B}  \propto \frac1{2G}\eps^\mu \eps^\nu (B_{\mu\nu}-C_{\mu\nu})$ defined
 by Eq. (1.7) of Ref. \cite{Sahoo:2021ctw}, see Appendix \ref{app:B}. Let us emphasize the presence of $O(\eta^3)$ and $O(\eta^5)$ contributions that are linked to tail effects. The existence of corresponding time-domain $O(1/t)$ tail terms in the $t\to +\infty$ limit was first pointed out in Ref. \cite{Blanchet:1992br} (see below Eq. (2.44) there).

In the MPM waveform (at the considered accuracy), the ${\cal C} \om \left(\ln \hat \om \right)^2 $ 
  contribution entirely comes from multiplying the frequency-domain tail
 logarithms (e.g., $\ln(2 b_0 |\om |)$ in Eq. \eqref{eq:U2_tail}) with a $\ln u$ term coming from the small-$u$ expansion of $K_0(u)$.
 As a consequence of this origin, and in view of the prefactor $\exp \left ( i \frac{2 G {\cal M \om}}{c^3} \ln \frac{ b_0 c \pinf}{b}\right) $ in Eq. \eqref{our_soft}, we found that, at our accuracy, we have the simple link
 \be \label{cC}
 { \cal C}^{\rm MPM}=   i \frac{2 G {\cal M }}{c^3}{ \cal B}^{\rm MPM} +O(G^3)\,.
 \ee
The result Eq. \eqref{cC} agrees (at the $G^2 \eta^5$ accuracy) with the corresponding result ${\cal C } \propto \frac1{2G}\eps^\mu \eps^\nu (F_{\mu\nu}-G_{\mu\nu})$ defined by  Eqs. (1.8), (1.9) of \cite{Sahoo:2021ctw}. The validity of the link Eq. \eqref{cC}
between the complicated expressions of $B_{\mu\nu}-C_{\mu\nu}$ and of $F_{\mu\nu}-G_{\mu\nu}$ (and thereby the
presence of a tail contribution, with a tail factor $ - 2G \sum_a p_a \cdot k= 2 G{\cal M }\om $ (in the c.m. frame)
 in $F_{\mu\nu}-G_{\mu\nu}$)
 is not a priori apparent when first looking at Eqs. (1.8), (1.9) of \cite{Sahoo:2021ctw}. 
However, it is shown in Eq. \eqref{BC_link} of Appendix \ref{app:B} that (in the two-body, conservative, c.m. frame case) the difference
${\mathcal C}-\frac{2iG {\mathcal M}}{c^3}{\mathcal B} $ is of order $O(G^3)$ when using the fact that $p_a'=p_a+O(G)$.

Let us emphasize in this respect that though some logarithms entering the 
 soft expansion are so related to the
 IR-divergence of the  gravitational ``Coulomb phase" (associated with the presence of the arbitrary time scale $b_0$), 
other logarithms  are associated to the logarithmic drift of the incoming and outgoing worldlines and  
are not directly linked to IR-divergences (containing arbitrary scales) in the MPM waveform, at least when the time-domain MPM
waveform is computed (as done here) from a solution of the equations of motion whose time origin is fixed at the moment of closest
approach of the two worldlines).
See in Appendix  \ref{app:B} the terms containing a tail factor  $GE$ versus the terms containing the coefficient $\Gamma$ of  the logarithmic drift of the worldlines.
The latter logarithms hide in the $K_0(u)$ and $K_1(u)$
 functions  (whose small-$u$ expansion contain impact-parameter-scaled logarithms: $\ln \hat\om=\ln u$). [Such logarithms are already present at the Newtonian approximation of $W(\om)$.] Let us also emphasize
 that the link  Eq. \eqref{cC} is only true at the $G^2$ level (in $W$). As shown in Appendix \ref{app:B}, the exact expressions Eqs. (1.8), (1.9) of 
 \cite{Sahoo:2021ctw} define quantities that have a rich structure which start at order  $G^3$ (in $h_{\mu\nu}$)
 but contain infinitely higher powers of $G$ (when expanding the differences $p'_a-p_a \sim G + G^2 + G^3 + \cdots$).

 As already mentioned, while Refs. \cite{Sahoo:2020ryf,Sahoo:2021ctw} obtained exact, universal expressions for  the first three soft-expansion coefficients $\cal A, \cal B, \cal C$ in terms of asymptotic dynamical data, there does not exist an analog, all-$G$-order expression for
the sub-sub-sub-leading soft term $+ {\mathcal D} \om \ln \hat \om $.  The presence of such a contribution (at order $G^1$ in $W$)
was first pointed out (in the high-energy limit) in Ref. \cite{Ciafaloni:2018uwe}, 
and was recently derived from the tree-level amplitude in Ref. \cite{DiVecchia:2023frv} (see Eq. (8.98) there, written in the high-energy limit).  Separately, Ref. \cite{Ghosh:2021bam} derived the leading-$G$-order expression of the corresponding $O(t^{-2})$ 
memories in terms of asymptotic dynamical data. The latter result only gives access to the $O(G^1)$ contribution to 
$ {\mathcal D}$, as given in Eq. \eqref{sahoo} in Appendix \ref{app:B}.

 We computed the sub-sub-sub-leading soft term $+ {\mathcal D} \om \ln \hat \om $ from our MPM results, which reach 
 the  loop-level ($G^1+G^2$) (at our usual $\eta^0+\eta^2+\eta^3+\eta^5$ PN-accuracy), and derived 
  explicit results both for the $G^1$ (tree level) and the $G^2$ (loop level) parts of
 ${\mathcal D}={\mathcal D}^{G^1} +{\mathcal D}^{G^2}$. The $O(G^1)$ part reads (in the
 equatorial plane, and even-in-$\phi$)
\bea
{\mathcal D}^{G^1}&=&\frac{GM^2 b}{p_\infty^2}\nu \frac{i }{2}\left[\sin(2\phi)  \right.\nonumber\\
&+&\left. \frac{1}{2} \eta^2 p_\infty^2 \left((3\nu - 4)\sin(2\phi)+\frac12   (3\nu - 1)\sin(4\phi)\right)\right].\nonumber\\
\eea
It does not contain $O(\eta^3)$ and $O(\eta^5)$ contributions. We checked that it agrees with the 
even-in-$\phi$ projection of the PN expansion of the leading-$G$-order result of Ref. \cite{Ghosh:2021bam}. See
Eqs.  \eqref{DG1sahoo_non_eq} and \eqref{Sigma_def} below in which one inserts $\epsilon_i=\bar m_i$
.

The MPM result for the $O(G^2)$ part  (which cannot be compared to existing soft results) can be decomposed into
${\mathcal D}^{G^2}={\mathcal D}^{G^2(\eta^0+\eta^2)}+{\mathcal D}^{G^2(\eta^3+\eta^5)}$
where the $G^2(\eta^3+\eta^5)$ part reads
${\mathcal D}^{G^2(\eta^3+\eta^5)}={\mathcal D}^{G^2}_{\rm tail}+{\mathcal D}^{G^2}_{QQ}+{\mathcal D}^{G^2}_{LQ}+ {\mathcal D}^{G^2}_{WQ}+{\mathcal D}^{G^2}_{\rm 2.5PN}$, as obtained  by using the corresponding expressions for $U_{ij}$, $U_{ijkl}$, $V_{ijk}$ given in Sec. II.  
The (equatorial-even-in-$\phi$) PN expansion of ${\mathcal D}^{G^2}$ then reads ($\gamma_E$ denoting Euler's constant)
\begin{widetext}
\bea
{\mathcal D}^{G^2(\eta^0+\eta^2)}&=& \frac{G^2M^3}{p_\infty^4}\frac{\pi \nu}{4} \left[  (\cos(2\phi) + 1)+\frac{1}{4}\eta^2 p_\infty^2 \left( 2(3\nu - 7) \cos(2\phi)  +  (3\nu - 1)\cos(4\phi)  +  (3\nu - 13)  \right) \right],\nonumber\\
{\mathcal D}^{G^2(\eta^3+\eta^5)}&=& \frac{\eta^3G^2M^3\nu}{p_\infty}\frac{1}{12}[ ( 24\gamma_E i + 6\pi - 35i)\cos(2\phi) +  ( 24\gamma_E i  + 6\pi - 11 i) ]\nonumber\\
&+&\frac{\eta^5G^2 M^3 p_\infty \nu}{40}\left[ \frac13 (360i\nu \gamma_E -  480i\gamma_E + 90\pi \nu - 873i \nu - 120\pi + 398i) \cos(2\phi) \right. \nonumber\\
&+&  \frac12 (120i \nu\gamma_E -  40\gamma_E i + 30\pi\nu - 421i\nu - 10\pi + 136i) \cos(4\phi) \nonumber\\
&+&\left. \frac16 (360i\nu\gamma_E -  840\gamma_Ei  + 90\pi\nu - 243i\nu - 210\pi + 1348i)\right]\,.
\eea
\end{widetext}

After having checked the compatibility of the soft expansion of the MPM waveform with known 
classical soft theorems, let us now turn to the soft expansion of the EFT waveform. This is most
easily discussed by considering our result above, Eqs.  \eqref{deltaW1}, \eqref{deltaW2}, for
the {\it time-domain  difference}, Eq. \eqref{diffW}, between the two waveforms.

While the $\ell=3^-$ time-domain multipole difference $\delta V_3(t)$, Eq. \eqref{deltaW2}, decays $\propto \frac1{t^3}$ as $t\to \pm \infty$, the mass-type multipole differences $\delta U_2(t)$ and $\delta U_4(t)$
 have a slower large-time decay, namely of order $\sim \frac{c^{\pm}_2}{t^2}$ 
 and $\sim \frac{c^{\pm}_4}{t^2}$,
 with different numerators $c^{\pm}_\ell$ as $t \to + \infty$ or $t \to - \infty$. 
Actually one finds that $c^{+}_\ell=-c^{-}_\ell$ for $\ell=2$ and $\ell=4$, and that they are respectively
proportional to $a_1 + a_2 + a_3 + a_4 + a_5=4\nu -\frac{16}{7}\nu^2 $ and 
$c_1 + c_2 +c_3+c_4+c_5+c_6=c_1+c_2=-24\nu^2  $.

The Fourier transform $f(\om)= \int dt e^{i \om t} f(t)$ of a function behaving,  
when $t \to \pm \infty$, as $\approx \frac{c^{\pm}}{t^2}$ is easily checked to 
behave, near $\om \to 0^+$, as
\be
f(\om) \approx - i (c^+ - c^-) \, \om \ln \om\,.
\ee

As a consequence, the soft limits ($\om \to 0^+$)
 of the (even-in-$\phi$) frequency-domain waveforms $W^{\rm EFT}(\om)$ and $W^{\rm MPM}(\om)$ 
 differ\footnote{Here, we talk about the EFT waveform referred to classical barred momenta. In view of Eq. \eqref{dWeta0}, the EFT waveform referred to incoming momenta would exhibit a much larger violation, starting at the leading  
 order of the soft expansion.} by 
 a $O(G^2/c^5)$    term $\propto \om \ln \om$, namely
 \be \label{dsoft}
 W^{\rm EFT}(\om) -W^{\rm MPM}(\om) \overset{\om \to 0}{=} (\delta {\mathcal D}_2 + \delta {\mathcal D}_4) \om \, \ln \om\,,
 \ee
 with 
 
 \bea \label{D2D4}
 \delta {\mathcal D}_2 &=& -i( c^{+}_2-c^{-}_2)=i \frac{(\eps \cdot v)^2}{ v_0}\frac{G^2M^3}{c^5}\left(-4\nu+\frac{16}{7}\nu^2 \right)\,, \nn\\
  \delta {\mathcal D}_4 &=&-i(c^{+}_4-c^{-}_4)\nonumber\\
&=& i \frac{ \eps_{ij} n_{kl}  [v^{ijkl}]^{\rm STF} }{v_0^3} \frac{ G^2M^3}{c^5} 
(2\nu^2)\,. \qquad
 \eea
The numerator of the last equation reads in general, 
\bea
&& \eps_{ij} n_{kl}  [v^{ijkl}]^{\rm STF}= \frac{2}{135} (\eps \cdot n)^2 v_0^4 
\\ \nonumber
&+& (\eps \cdot v)^2 \left((n \cdot v)^2 - \frac{19}{189} v_0^2  \right) - \frac{4}{7} (\eps \cdot n) (\eps \cdot v) (n \cdot v) v_0^2,
\eea
and takes the following value on the equatorial plane
\beq
 \eps_{ij} n_{kl}  [v^{ijkl}]^{\rm STF} =\frac{v_0^4}{112} (-3 + 7\cos(4\phi) + 4\cos(2\phi))\,.
\eeq

The result \eqref{dsoft} characterises
 a difference, in the PN-expanded EFT waveform, with respect to the  soft behavior
 of the MPM waveform. 
 This difference is sub-sub-sub-leading
 in the PM-expanded soft expansion of the Fourier-domain waveform, Eqs. \eqref{soft0}, \eqref{our_soft}.
 The (corrected) values, Eqs. \eqref{D2D4}, of the soft-expansion coefficients in Eq. \eqref{dsoft}
  agree with Eq. (5.25) in Ref. \cite{Georgoudis:2023eke}.

\section{Conclusions}

A first conclusion of our work is that computing the integrated one-loop waveform by systematically
using the incoming momenta in its definition leads to large (physically meaningful) differences, which start at the Newtonian level,
$O\left(\frac{G^2}{c^0} \right)$,
with the corresponding MPM waveform. However, these differences are much reduced when computing
 the integrated one-loop waveform by systematically using instead (both in the measure and in the integrand)
 the classical barred momenta ${\bar p}_a$.
However, even when doing so there remain many differences, of order $O\left(\frac{G^2}{c^5} \right)$,
in $W = \frac{2 i}{\kappa } {\mathcal M}^{\rm integrated}(\om, \theta,\phi)$
 between the recent one-loop EFT computations
of the gravitational bremsstrahlung waveform \cite{Brandhuber:2023hhy,Herderschee:2023fxh,Georgoudis:2023lgf}
and the corresponding result of the Multipolar-Post-Minkowskian (MPM) formalism. 
As discussed in Section \ref{secsoft} (and Appendix \ref{app:B})
the remaining $O\left(\frac{G^2}{c^5} \right)$ time-domain
waveform difference corresponds to a frequency-domain difference that is at the sub-sub-sub-leading
order in the soft expansion. Had we referred the EFT waveform to the incoming momenta, the EFT-MPM
difference would start at the Newtonian level, and at the leading-order in the soft expansion.

The MPM formalism has been developed over the
last 35 years, with many internal and external checks and recomputations,
and has become the most accurate analytical approach for perturbatively computing 
the generation of gravitational waves (GW) by generic sources. It seems then  probable that the EFT$-$MPM differences come from
the EFT side, especially for those arising at the linear-in-mass-ratio level in the quadrupolar GW tail.

Let us also mention that the cubically nonlinear contributions to the radiative quadrupole
have been recently recomputed by using new variants of the MPM formalism \cite{Trestini:2023wwg,Blanchet:2023sbv}.

We have emphasized that all the ($\bar p$-frame) differences we have detected~\footnote{We recall that we focussed on the multipolar
contributions $2^+$, $3^-$ and $4^+$. There could also be differences in the unexplored multipolar contributions 
$2^-$, and $3^+$.} are
linked to time-asymmetric effects: some (in $\delta U_2$) contain contributions directly linked to time-odd 
radiation-reaction effects in the worldline motion; others are linked to various (linear and non-linear) retarded 
contributions to the waveform. In particular, we have highlighted in Fig. 1 and Fig. 2 the couplings (and
the diagrams) behind various $O\left(\frac{G^2}{c^5} \right)$ contributions.  Fig. 1 displays the effect
of the coupling of gravitational radiation with a time-independent part of the metric generated by
the system (which can be the  $\frac{G \cal M}{r}$ potential for tail effects, or the angular-momentum-generated,
 $\frac{G  L}{r^2}$, part of the metric for the terms $U_{ij}^{\rm LQ}$, Eq. \eqref{U2LQ}, or  $V_{ij}^{\rm LQ}$, Eq. 
 \eqref{V3tail}. Fig. 2 illustrates the effect of the coupling of (quadrupolar) gravitational radiation on the radiative 
 (quadrupolar) part of the metric, corresponding to a QQQ term in the action. The corresponding (MPM predicted)
 $U_2^{\rm QQ}$, $V_3^{\rm QQ}$ and $U_4^{\rm QQ}$ contributions to the waveform are displayed
 in Eqs. \eqref{U2QQ}, \eqref{V3tail} and \eqref{U4tail} respectively.

While this work was being written up, we learned of  Ref. \cite{Caron-Huot:2023vxl} in which it is notably argued that
some of the master integrals computed in \cite{Brandhuber:2023hhy,Herderschee:2023fxh,Georgoudis:2023lgf}
need to be completed. Our results give, at the perturbative order $\frac{G^2}{c^5}$, a precise measure of
any putative, future completion of the one-loop waveform.

\appendix

\section{1PN corrections to the MPM radiative multipole moments $V_3$ and $U_4$}
\label{1PN_corr_V3_U4}

In this appendix we  display the equatorial values of the 1PN contributions $O(G\eta^2)$ and   $O(G^2\eta^2)$ to the MPM radiative multipole moments $V_3$ and $U_4$, as indicated in Eqs. \eqref{V3_struc} and \eqref{U4_struc} above. The 1PN corrections to $V_3$ at $O(G)$ and $O(G^2)$  are given by
\bea
\label{V3_1PN}
V_3^{G {\eta^2}}&=& -p_\infty^3 GM^2\nu u \left[A_{30}^{G\eta^2}K_0(u)+A_{31}^{G\eta^2}K_1(u)\right]\,,\nonumber\\
V_3^{G^2 {\eta^2}}&=& -\frac{\nu G^2M^3 p_\infty \pi}{b} \left[A_{30}^{G^2\eta^2}K_0(u)+A_{31}^{G^2\eta^2}K_1(u)\right.\nonumber\\
&+& \left. B_3^{G^2 {\eta^2}}e^{-u}\right]\,,
\eea
where
\bea
A_{30}^{G\eta^2}&=& \left(-\frac{5}{6} u\nu^2-\frac{1}{6} u\nu+\frac{1}{10} u\right)\cos(2\phi)\nonumber\\
&+& \left(\frac{13}{27}\nu^2 u^2-\frac{5}{18}\nu^2-\frac{77}{54} u^2\nu\right. \nonumber\\
&-&\left. \frac{19}{18}\nu+\frac{52}{135} u^2+\frac{13}{36}\right)i\sin(2\phi)\nonumber\\
A_{31}^{G\eta^2}&=& \left(-\frac{4}{9}+\frac{52}{135} u^2+\frac{13}{27}\nu^2 u^2\right.\nonumber\\
&-&\left. \frac{5}{9}\nu^2-\frac{77}{54} u^2\nu+\frac{25}{18}\nu\right)\cos(2\phi)\nonumber\\
&+&\left(\frac{79}{270} u-\frac{95}{108} u\nu-\frac{16}{27} u\nu^2\right)i\sin(2\phi)\nonumber\\
A_{30}^{G^2\eta^2}&=& \left(-\frac{1}{6} u^3\nu^2+\frac{3}{10} u^3-\frac{11}{12} u^3\nu \right)\cos(2\phi)\nonumber\\
&+&\left(-\frac{17}{18} u^2\nu-\frac{77}{108} u^4\nu+\frac{11}{36} u^2-\frac{1}{72} \nu^2 u^2\right.\nonumber\\
&+&\left. \frac{13}{54} u^4\nu^2+\frac{26}{135} u^4\right)i \sin(2\phi)\nonumber\\
A_{31}^{G^2\eta^2}&=& \left(\frac{1}{36} u^2-\frac{1}{36}\nu^2 u^2-\frac{5}{36} u^2\nu+\frac{13}{54} u^4\nu^2\right.\nonumber\\
&-& \left. \frac{77}{108} u^4\nu+\frac{26}{135} u^4\right)\cos(2\phi)\nonumber\\
&+&\left(-\frac{275}{216} u^3\nu+\frac{107}{270} u^3-\frac{5}{108} u^3\nu^2\right)i\sin(2\phi)\nonumber\\
B_3^{G^2 {\eta^2}}&=& \frac12(u^2+u)(3\nu-1) 
i\sin(2\phi)\nonumber\\
&+&  \frac12 (1+u+u^2)\left(3\nu -1 \right)\cos(2\phi) \,.
\eea

Similarly, the 1PN corrections to $U_4$ at $O(G)$ and $O(G^2)$  are given by
\bea
\label{U4_1PN}
U_4^{G {\eta^2}}&=&-\frac{27}{154}p_\infty^3M^2\nu G \eta^2\left[A_{40}^{G\eta^2}K_0(u)+A_{41}^{G\eta^2}K_1(u)\right.\nonumber\\
& +& \left. B^{G\eta^2}e^{-u} \right]\,, \nonumber\\
\nonumber\\
U_4^{G^2 {\eta^2}}&=& \eta^2\left[A_{40}^{G^2\eta^2}K_0(u)+A_{41}^{G^2\eta^2}K_1(u)\right.\nonumber\\ 
&+&\left. B^{G^2\eta^2}e^{-u} \right] \,,
\eea
where
\begin{widetext}
\bea
A_{40}^{G\eta^2}&=&  
\left( \left(\nu ^2-\frac{136 \nu }{81}+\frac{709}{1620}\right) u^3+\left(-\frac{\nu ^2}{18}-\frac{19 \nu }{54}+\frac{1}{20}\right) u\right) i\sin (2 \phi ) \nonumber\\
&&
+\left(-\frac{7}{27} 
   \left(\nu ^2+\frac{131 \nu }{12}-\frac{361}{120}\right) u^3-\frac{7}{27}  \left(\frac{3 \nu ^2}{4}+\frac{19 \nu }{4}-\frac{27}{40}\right) u\right) i\sin (4 \phi )\nonumber\\
&&
+\left(\left(\frac{41
   \nu ^2}{54}-\frac{313 \nu }{108}+\frac{901}{1080}\right) u^2+\frac{\nu ^2}{9}-\frac{19 \nu }{6}+1\right) \cos (2 \phi )\nonumber\\
&&
+\left(\left(\frac{56 \nu ^2}{27}-\frac{413 \nu
   }{162}+\frac{301}{540}\right) u^4+\left(-\frac{35 \nu ^2}{54}-\frac{497 \nu }{144}+\frac{3983}{4320}\right) u^2+\frac{7 \nu ^2}{36}-\frac{133 \nu }{24}+\frac{7}{4}\right) \cos (4 \phi
   )\nonumber\\
&& 
+\left(-\frac{23 \nu ^2}{27}+\frac{1039 \nu }{432}-\frac{607}{864}\right) u^2-\frac{\nu ^2}{12}+\frac{19 \nu }{8}-\frac{3}{4}\nonumber\\
A_{41}^{G\eta^2}&=&  \left(\left(\nu ^2-\frac{136 \nu }{81}+\frac{709}{1620}\right) u^3+\left(\frac{59 \nu ^2}{54}-\frac{745 \nu }{108}+\frac{373}{180}\right) u\right) \cos (2 \phi ) \nonumber\\
&&
+\left(\left(\frac{7 \nu
   ^2}{9}-\frac{665 \nu }{162}+\frac{343}{324}\right) u^3+\left(\frac{35 \nu ^2}{216}-\frac{2149 \nu }{432}+\frac{91}{80}\right) u\right) \cos (4 \phi )\nonumber\\
&&
+\left(\left(\frac{34  \nu
   ^2}{27}-\frac{1211 \nu }{324}+\frac{853 }{810}\right) u^2+\frac{2  \nu ^2}{9}+\frac{20 \nu }{9}-\frac{23 }{27}\right) i\sin (2 \phi )\nonumber\\
&&
+\left(\left(\frac{56 \nu
   ^2}{27}-\frac{413 \nu }{162}+\frac{301 }{540}\right) u^4+\left(\frac{616 }{405}-\frac{3773  \nu }{648}\right) u^2+\frac{7 \nu ^2}{9}+\frac{70 \nu }{9}-\frac{161 }{54}\right)
   i\sin (4 \phi )\nonumber\\
&& 
+\left(-\frac{77 \nu ^2}{72}+\frac{99 \nu }{16}-\frac{275}{144}\right) u\,.
\eea
Finally
\bea
A_{40}^{G^2\eta^2}&=&-\frac{9 \pi  G^2 M^3 \nu  p_\infty u}{44 b}\left[
\left(\left(\frac{3 \nu ^2}{7}-\frac{136 \nu }{189}+\frac{709}{3780}\right) u^3+\left(\frac{31 \nu ^2}{84}-\frac{92\nu }{63}+\frac{29}{70}\right) u\right) i\sin (2 \phi
   )\right.\nonumber\\
&&
+\left(\left(\frac{ \nu ^2}{2}-\frac{13  \nu }{4}+\frac{1021}{1080}\right) u^3+\left(\frac{31 \nu ^2}{24}-\frac{46 \nu }{9}+\frac{29 }{20}\right) u\right) i \sin (4 \phi
   )\nonumber\\
&&
+\left(\left(\frac{\nu ^2}{2}-\frac{197 \nu }{108}+\frac{149}{280}\right) u^2+\frac{13 \nu ^2}{42}-\frac{281 \nu }{126}+\frac{29}{42}\right) \cos (2 \phi )\nonumber\\
&&
+\left(\left(\frac{8 \nu
   ^2}{9}-\frac{59 \nu }{54}+\frac{43}{180}\right) u^4+\left(\frac{17 \nu ^2}{18}-\frac{2399 \nu }{432}+\frac{2329}{1440}\right) u^2+\frac{13 \nu ^2}{24}-\frac{281 \nu
   }{72}+\frac{29}{24}\right) \cos (4 \phi )\nonumber\\
&&\left.
+\left(-\frac{23 \nu ^2}{63}+\frac{1039 \nu }{1008}-\frac{607}{2016}\right) u^2-\frac{13 \nu ^2}{56}+\frac{281 \nu }{168}-\frac{29}{56}
\right]\,,  \nonumber\\
A_{41}^{G^2\eta^2}&=&-\frac{9 \pi  G^2 M^3 \nu  p_\infty u}{44 b}\left[
\left(\left(\frac{3 \nu ^2}{7}-\frac{136 \nu }{189}+\frac{709}{3780}\right) u^3+\left(\frac{239 \nu ^2}{252}-\frac{3445 \nu }{756}+\frac{431}{315}\right) u\right) \cos (2 \phi
   )\right.\nonumber\\
&&
+\left(\left(\frac{17 \nu ^2}{18}-\frac{205 \nu }{54}+\frac{115}{108}\right) u^3+\left(\frac{329 \nu ^2}{144}-\frac{4111 \nu }{432}+\frac{973}{360}\right) u\right) \cos (4 \phi
   )\nonumber\\
&&
+\left(\left(\frac{5  \nu ^2}{7}-\frac{1651  \nu }{756}+\frac{169 }{270}\right) u^2+\frac{13 \nu ^2}{21}-\frac{50  \nu }{63}+\frac{10 }{63}\right) i\sin (2 \phi
   )\nonumber\\
&&
+\left(\left(\frac{8  \nu ^2}{9}-\frac{59  \nu }{54}+\frac{43 }{180}\right) u^4+\left(\frac{55  \nu ^2}{36}-\frac{1639 \nu }{216}+\frac{1177 }{540}\right) u^2+\frac{13  \nu^2}{6}-\frac{25 \nu }{9}+\frac{5}{9}\right) i\sin (4 \phi )\nonumber\\
&&\left.
+\left(-\frac{209 \nu ^2}{336}+\frac{3223 \nu }{1008}-\frac{55}{56}\right) u
\right] \,,  \nonumber\\
B^{G^2\eta^2}&=&-\frac{3 \pi  G^2 M^3 \left(\nu -\frac{1}{3}\right) \nu  p_\infty}{2 b u}\left[
\left(\frac{ u^3}{14}+\frac{3 u^2}{14}+\frac{3 u}{7}+\frac{3}{7}\right) i\sin (2 \phi )+\left(\frac{u^3}{14}+\frac{3 u^2}{14}+\frac{3 u}{14}\right) \cos (2 \phi )\right.\nonumber\\
&&\left.
+\left(\frac{u^4}{2}+ u^3+\frac{3 u^2}{2}+\frac{3 u}{2}+\frac{3}{2}\right)i \sin (4 \phi )+\left(\frac{u^4}{2}+u^3+\frac{3 u^2}{2}+\frac{3 u}{2}\right) \cos (4 \phi )
\right] \,.
\eea
\end{widetext}

\section{Exact expressions for the soft-expansion coefficients ${\mathcal A}$, ${\mathcal B}$, ${\mathcal C}$ of Eq. \eqref{our_soft} in the c.m. two-body  conservative case}
\label{app:B}

Following Refs. \cite{Sahoo:2020ryf,Sahoo:2021ctw} we list below the exact expressions the soft-expansion coefficients ${\mathcal A}$, ${\mathcal B}$, ${\mathcal C}$ of Eq. \eqref{our_soft}, limiting ourselves to the case of interest here, namely $m=n=2$ particles in the initial and final state, conservative scattering, c.m. frame. We  use a standard notation, denoting as $p_1$ and $p_2$ the incoming momenta of the two bodies, and as $p_1'$ and $p_2'$ the corresponding outgoing momenta.
The soft-expansion coefficients of Eq. \eqref{our_soft} read, in terms of the notation of Refs. \cite{Sahoo:2020ryf,Sahoo:2021ctw}:
\bea
{\mathcal A} &=& \frac{ r}{2G}  i  A_{\mu\nu}\epsilon^\mu \epsilon^\nu \,, \nonumber\\
{\mathcal B}  &=& -\frac{ r}{2G}(B_{\mu\nu}-C_{\mu\nu})\epsilon^\mu \epsilon^\nu \,,  \nonumber\\
{\mathcal C}   &=& \frac{ r}{2G} \frac{i}{2}(F_{\mu\nu}-G_{\mu\nu})\epsilon^\mu \epsilon^\nu\,, 
\eea
with the explicit expressions of the tensors $A_{\mu\nu}$, $B_{\mu\nu}$, $C_{\mu\nu}$, $F_{\mu\nu}$, $G_{\mu\nu}$ given in Eqs. (1.6) - (1.9) of Ref. \cite{Sahoo:2021ctw}.
It is convenient to define the scalar quantity 
\beq
\Gamma =\frac{3\gamma-2\gamma^3}{(\gamma^2-1)^{3/2}}\,,
\eeq
measuring the relative logarithmic drift of the two worldlines, see e.g.,  Eq. (4.9) of Ref. \cite{Bini:2018ywr}, 
as well as (denoting here $n^\mu\equiv k^\mu/\omega$) the vectors
\bea
{\mathcal D}^\mu&=& \frac{p_1^\mu}{(n\cdot p_1)}-\frac{p_2^\mu}{(n\cdot p_2)}=\frac{[p_1\wedge p_2]^{\mu\alpha}n_\alpha}{(n\cdot p_1)(n\cdot p_2)} \,,\nonumber\\
{\mathcal D'}^\mu&=& \frac{p_1'^\mu}{(n\cdot p_1')}-\frac{p_2'{}^\mu}{(n\cdot p_2')}=\frac{[p_1'\wedge p_2']^{\mu\alpha}n_\alpha}{(n\cdot p_1')(n\cdot p_2')} \,,
\eea
associated to the (one-copy) photon memory.
Using the compact notation  $a \cdot b =(ab)$ for the scalar product of two generic vectors $a$ and $b$,  we find 
\begin{widetext}
\bea
{\mathcal A} 
&=&   i  \left[ - \frac{(\epsilon p_1')^2}{(n p_1')}- \frac{(\epsilon p_2')^2}{(n p_2')} +
\frac{(\epsilon   p_1)^2}{(n  p_1)}
+\frac{(\epsilon  p_2)^2}{(n  p_2)}\right] \,,\nonumber\\
{\mathcal B}&=& 2iG E {\mathcal A}  -\Gamma G[(np_1' )(np_2' )(\epsilon {\mathcal D}')^2 + (np_1 )(np_2 ) (\epsilon {\mathcal D} )^2 ]\,,\nonumber\\
{\mathcal C}&=&-2G^2 E^2 {\mathcal A}-i G^2 E \Gamma\left[\left(2+\frac{\Gamma}{2}\right) (np_1' )(np_2' )(\epsilon {\mathcal D}')^2
+\left(2-\frac{\Gamma}{2}\right)(np_1 )(np_2 ) (\epsilon {\mathcal D} )^2  \right]
\eea
implying the exact link
\beq
\label{BC_link}
{\mathcal C}=  2i  G  E  {\mathcal B} -\frac{i}{2}  G^2 E \Gamma^2 [(np_1' )(np_2' )(\epsilon {\mathcal D}')^2 - (np_1 )(np_2 ) (\epsilon {\mathcal D} )^2 ]+ 2G^2E^2{\mathcal A}\,.
\eeq
\end{widetext}
Note the presence in the equations above
 of the tail factor $2GE$, with $E=E^{\rm tot}_{\rm c.m.}={\mathcal M}c^2$,  as well as of the logarithmic drift factor $\Gamma$. 
 
Finally, applying the leading-$G$-order results of Ref. \cite{Ghosh:2021bam} to our case ($m=n=2$ particles in the initial and final state) yields the following ($O(G)$-accurate) expression for ${\mathcal D}$
\beq \label{sahoo}
{\mathcal D}=-i2G\Gamma [\epsilon  p_1 p_2  n] \left( \frac{ [\epsilon  b^-_1 p_1  n]}{( p_1n )} -\frac{ [\epsilon  b^-_2 p_2  n]}{(p_2n  )}\right) +O(G^2)\,.
\eeq
Here, $p_1$ and $p_2$ are the incoming momenta, $b^-_1$ and $b^-_2$ the incoming
vectorial impact parameters, and we have introduced the compact notation
\bea
[abcd]&=&(ab)(cd)-(ac)(bd)\nonumber\\
&=& \frac12  [a\wedge d]^{\mu\nu}[b\wedge c]_{\mu\nu}\nonumber\\
&=& -[acbd]=-[dbca]=[cdab]\,.
\eea
At its indicated $O(G^1)$ accuracy, Eq. \eqref{sahoo} can be re-expressed by replacing $p_1$ and $p_2$ by
the barred momenta ${\bar p}_1$ and ${\bar p}_2$, and the incoming impact parameters by the corresponding eikonal-type ones
$b_1$, $b_2$.  It is the latter expression that we call ${\mathcal D}^{G^1}$ below.
\begin{widetext}
In our notation, and with  
$X_{12}\equiv \frac{m_1-m_2}{m_1+m_2}$, the just defined quantity  ${\mathcal D}^{G^1}$ reads
\bea
\label{DG1sahoo_non_eq}
{\mathcal D}^{G^1}&=& \frac{2im_1m_2 Gb \gamma(2\gamma^2-3)}{(\gamma^2-1)}\epsilon_y \left(-\epsilon_x+\Sigma \epsilon_y\right)\,,
\eea
with
\bea
\label{Sigma_def}
\Sigma&=& 
\frac{\sin\theta\cos\phi\sqrt{\gamma^2-1}[X_{12}-\sin\theta\sin\phi  \nu \sqrt{\gamma^2-1}]}{\sin\theta \sin\phi \sqrt{\gamma^2-1} X_{12}+ \nu(\gamma^2-1)(\cos^2\theta\sin^2\phi+\cos^2\phi)+\gamma-2\nu(\gamma-1)}\,.
\eea
\end{widetext}
Let us recall that we generally use ${\boldsymbol\epsilon}=\bar {\mathbf m}$,  with   explicit  components  
\bea
\bar m_x &=&\frac{1}{\sqrt{2}}  (\cos(\theta)\cos(\phi) + i\sin(\phi) ) \,, \nonumber\\
\bar m_y &=& \frac{1}{\sqrt{2}}(\cos(\theta)\sin(\phi) - i\cos(\phi))\,,  \nonumber\\
\bar m_z &=& -\frac{1}{\sqrt{2}} \sin(\theta)\,.
\eea

The high-energy limit ($\gamma\to \infty$, $m_1\to 0$,  $m_2\to 0$ with $E \approx 2 P_{\rm c.m.} \approx \sqrt{2 \gamma m_1 m_2}$ finite) of ${\mathcal D}^{G^1}$, Eq. \eqref{sahoo},  can be written  in terms of the
c.m. spatial vectors ${\bf n}$, ${\bf b}\equiv b e_x$, ${\bf P} \equiv {\bar P}_{\rm c.m.} e_y \equiv {\bar  P}_{\rm c.m.} {\hat {\bf P}} \simeq { P}_{\rm c.m.} {\hat {\bf P}}$ (we henceforth neglect the $O(G^2)$ difference between ${\bar  P}_{\rm c.m.}$ and ${  P}_{\rm c.m.}$),
${\bf P}_{\perp {\bf n}}\equiv {\bf P} - ({\bf n} \cdot  {\bf P}) {\bf n}$, and  ${\bf n} \times {\bf P}={\bf n} \times {\bf P}_{\perp {\bf n}} $, and of a general transverse, null, spatial polarization vector $\boldsymbol{ \epsilon}$ as
(denoting $( {\bf a}, {\bf b}, {\bf c}) \equiv \epsilon_{ijk} a^i b^j c^k$)
\beq 
\label{DG11}
{\mathcal D}^{G^1,\rm HE}=i 8 G \frac{({\bf n}, {\bf b}, {\hat {\bf P}})}{1- ({\bf n} \cdot {\hat {\bf P}})^2} (\boldsymbol{ \epsilon} \cdot {\bf P}_{\perp {\bf n}}) (\boldsymbol{ \epsilon} \cdot{\bf n} \times {\bf P}_{\perp {\bf n}}) \,.
\eeq
This vanishes in the equatorial plane (because of the factor $({\bf n}, {\bf b}, {\hat {\bf P}})= b \, {\bf n} \cdot e_z= b \cos \theta $) and
  remains finite (and actually vanishes) in the collinear limit where ${\bf n}$ becomes parallel to ${\bf P}= P_{\rm c.m.} {\hat {\bf P}}$. Indeed, the
  denominator $1- ({\bf n} \cdot {\hat {\bf P}})^2$ is equal to $ ({\bf n} \times {\hat {\bf P}})^2 = {\bf P}_{\perp {\bf n}}^2/P_{\rm c.m.}^2$, so that we can rewrite ${\mathcal D}^{G^1,\rm HE}$, Eq. \eqref{DG11}, as
  \bea \label{DG12}
 {\mathcal D}^{G^1,\rm HE} &=&i 8 G P_{\rm c.m.}^2 ({\bf n}, {\bf b}, {\hat {\bf P}}) \epsilon_X \epsilon_Y \nn \\
 &=&i 8 G b  P_{\rm c.m.}^2 ({\bf n} \cdot e_z)\,  \epsilon_X \epsilon_Y \,,
 \eea
 involving  the product $\epsilon_X \epsilon_Y$ of the components of the polarization vector $\boldsymbol{ \epsilon}$
 along the two orthogonal unit vectors $e_X=  {\bf P}_{\perp {\bf n}}/|{\bf P}_{\perp {\bf n}}|$ and $e_Y=  ({\bf n} \times {\bf P})/|{\bf n} \times {\bf P}|$,
which form an orthonormal triad with ${\bf n}$. The expression  Eq. \eqref{DG12} shows that the (real) $O(1/t^2)$ memory waveform $\Delta c_{\mu \nu} = c_{\mu \nu}^{+}- c_{\mu \nu}^{-}$ 
(linked to  $\frac{R}{4G}h_{\mu \nu}(t \to \pm \infty)\sim c_{\mu \nu}^{\pm}/t^2$) giving rise to $ {\mathcal D}= - i \Delta c_{\mu \nu} \epsilon^\mu \epsilon^\nu$ becomes, in the high-energy limit, completely linearly polarized in the $\Delta c_{XY}$ (cross) direction,
with 
\bea
\Delta c_{XY}&=& - 4   G b  P_{\rm c.m.}^2 ({\bf n} \cdot e_z) \nonumber\\
&=& -  G b E^2 ({\bf n} \cdot e_z) \nonumber\\
&=& -  G b E^2  \cos \theta\,.
\eea
We have checked that the $e_X e_Y$ cross polarization coincides with the $e_{\theta} e_{\phi}$ cross polarization considered in Ref. \cite{DiVecchia:2023frv}.
[More precisely, we have $e_X= - e_{\theta}$, $e_Y= - e_{\phi}$ in terms of the (purely spatial) polarization vectors defined
in  Eq. (E.19) of \cite{DiVecchia:2023frv}.] We have also checked that  the ratio of the high-energy limits of  
${\mathcal D}_{\rm HE}$ and ${\mathcal A}_{\rm HE}$ is equal to
\bea \label{DbyA}
\frac{{\mathcal D}_{\rm HE}}{{\mathcal A}_{\rm HE}}&=&\frac{b^2}{4}(\cos^2\phi+\sin^2\phi\cos^2\theta)\nonumber\\
&=&\frac{b^2}{4}  |{\bf n} \times {\hat{\bf P}}|^2\,,
\eea
independently of the choice of the null complex polarization vector $\boldsymbol{ \epsilon}$ (indeed the latter ratio
is invariant under $\boldsymbol{ \epsilon} \to e^{i \alpha}\boldsymbol{ \epsilon}$). The result Eq. \eqref{DbyA} agrees
with Eq. (8.98) of \cite{DiVecchia:2023frv}.

\section*{Acknowledgements}

We thank Z. Bern, L. Blanchet, S. Caron-Huot, G. Chen, P. Damgaard,  S. de Angelis, P. di Vecchia, C. Heissenberg, A. Herderschee, R. Roiban,  R. Russo, A. Sen and F. Teng 
for several informative exchanges. The present research was
partially supported by the 2021 Balzan Prize for Gravitation: Physical and Astrophysical Aspects, awarded
to T. Damour. D.B.  
acknowledges sponsorship of the Italian Gruppo Nazionale per la Fisica Matematica (GNFM)
of the Istituto Nazionale di Alta Matematica (INDAM), 
as well as the hospitality and the highly stimulating environment of the Institut des Hautes Etudes Scientifiques.


\end{document}